\documentclass[fleqn,usenatbib]{mnras}

\usepackage{newtxtext,newtxmath}

\usepackage[T1]{fontenc}

\DeclareRobustCommand{\VAN}[3]{#2}
\let\VANthebibliography\thebibliography
\def\thebibliography{\DeclareRobustCommand{\VAN}[3]{##3}\VANthebibliography}

\usepackage{graphicx}	
\usepackage{amsmath}	
\usepackage{appendix}

\title[]{Pickles on FIRE: The 3D Shape Evolution of Simulated Milky Way–Mass Galaxies}

\author[Xia et al.]{
Luke Y. Xia,$^{1}$\thanks{E-mail: lyxia@uci.edu}
Courtney Klein,$^{1}$
Jenny D. Wang,$^{1}$
James Bullock,$^{1,2}$
Michael Boylan-Kolchin,$^{3}$
\newauthor Vincent Caudillo,$^{1}$
Jorge Moreno,$^{4,5}$
Francisco J. Mercado,$^{4,6}$
and Robert Feldmann$^{7}$
\\
$^{1}$Department of Physics and Astronomy, University of California Irvine, Irvine, CA 92697, USA\\
$^{2}$Department of Physics and Astronomy, University of Southern California, Los Angeles, CA 90089, USA\\
$^{3}$Department of Astronomy, The University of Texas at Austin, Austin, TX 78712, USA\\
$^{4}$Department of Physics and Astronomy, Pomona College, Claremont, CA 91711, USA\\
$^{5}$Carnegie Observatories, Pasadena, CA 91101, USA\\
$^{6}$TAPIR, California Institute of Technology, Pasadena, CA 91125, USA\\
$^{7}$Institute for Computational Science, University of Zurich, Zurich CH-8057, Switzerland
}

\date{Accepted XXX. Received YYY; in original form ZZZ}

\pubyear{\the\year{}}

\begin{document}
\label{firstpage}
\pagerange{\pageref{firstpage}--\pageref{lastpage}}
\maketitle

\begin{abstract}
We use reduced-mass eigentensors to quantify the 3D ellipsoidal shape evolution of thirteen Milky Way-mass galaxies simulated using zoom simulations with FIRE-2 physics; all but one form disks at $z=0$. 
We find that all of our Milky Way progenitors go through phases when they are elongated.  They often oscillate between spheroidal and elongated shapes in the early Universe over billion-year timescales, with $\sim 25-45\%$ of the population having elongated luminosity-weighted shapes at any given time at $z = 0.5-8.5$. In contrast, all stellar populations in our $z=0$ Milky Way analogs are symmetric about their minor axes at $z=0$, even though the old and intermediate-age stellar populations were often arranged in the shape of elongated pickles or triaxial spheroids at the time they formed meaning these populations changed shape significantly over time. During their transient elongated phases, our galaxies have anisotropic velocity dispersion ellipsoids directed along their spatial major axis; however, their shapes {\em do not} correlate with their dark matter fraction nor with the shapes and orientations of their underlying dark matter halos. We find that when treated as a population, the fraction of our galaxy progenitors that are elongated at $z>0.5$ is roughly consistent with what is observed for systems of the same mass and redshift. Our results suggest that observed elongated galaxies seen in the early Universe with JWST and HST are not stable structures, but rather transitory phases that are nevertheless statistically common. Some of these observed objects may evolve into Milky Way-like galaxies at $z=0$.
\end{abstract}




\section{Introduction}\label{sec:intro}

The origin and evolution of the three-dimensional structure of galaxies is among the core questions in galactic astrophysics \citep{Hubble1926,Sandage1970,Binney1981, Conselice2014,Favaro2025}. Galaxy shapes are fundamental to our understanding of the Universe and offer important tests for modern theories of galaxy formation \citep{White1978,Blumenthal1984,Springel2005,Somerville2015,Arora2025, 2022MNRAS.516.2389V}.

The Milky Way can serve as a powerful window into galactic structure \citep{Juric2008,Haywood2013,Bovy2016}.  We know, for example, that the youngest Milky Way stars reside within a flattened disk and that older stars are more isotropically distributed. However, relating the shape distribution of the oldest stars today to the shape of the Galaxy at the time when those stars were forming is non-trivial owing to the possibility of dynamical phase mixing and relaxation \citep{Merritt1996}. For this reason, a detailed understanding of the assembly of Milky Way–type galaxies must be informed by theoretical models combined with empirical studies of Milky Way–like progenitors in the distant Universe.

For the vast majority of galaxies, we must observationally infer information about their 3D shapes using their 2D projections on the sky. Under the assumption that galaxies are well-approximated by triaxial ellipsoids \citep{Binney1978}, it is possible to infer their 3D properties statistically using observed 2D axis ratios of galaxy populations \citep{Sandage1970,Fasano1993,Ryden2004}. Under the ellipsoidal approximation, we can think of galaxies as falling within three broad categories: spheroidal (all three axes roughly equal, like an orange), disky (flattened in one direction, like a pancake), or elongated (flattened in two directions, like a pickle). 

The question of the physical origin of elongated galaxy shapes is particularly vibrant. Mergers with specific configurations can play an important role.  For example, \citet{Lokas14} explained the prolate rotation observed in the  Local Group galaxy Andromeda II as a merger between two disky dwarf galaxies.
\citet{2017A&A...606A..62T} show that polar major mergers of disks can generally produce prolate shapes with prolate rotation. Mergers are often directed along large-scale filaments.  Indeed, \citet{Tomassetti2016} find that in their simulations, galaxy elongation is supported by an anisotropic velocity dispersion that originates from the assembly of the galaxy along a dominant large-scale filament. The fact that galaxy formation is fed along streams of the cosmic web is well-studied and known to be especially important at higher redshifts \citep[see][and references therein]{Dekel09,Stewart11,Tacchella16b}. It is known observationally that galaxy properties correlate with their spin orientations with respect to filaments \citep{Barsanti2022}.

By examining projected axis distributions, observational studies have concluded that the majority of massive quenched galaxies in the local Universe are spheroidal and that massive star-forming galaxies at $z=0$ are usually disks like the Milky Way
\citep{Lambas1992,Padilla2008,deNicola2022}.
At lower masses, the situation is somewhat more ambiguous: most studies suggest that low-mass galaxies at $z=0$ are a mixture of oblate disks and spheroids \citep{Ichikawa89,Binggeli95,Sung98,Roychowdhury2013,Putko2019,KF2020,Rong2020}, while others have found that a substantial population of local galaxies in this regime may be elongated or prolate in shape \citep{Vincent2005,Burkert2017}.

At higher redshifts, galaxies are known to exhibit more irregular, clumpy morphologies than we see in the local Universe \citep[e.g.][]{Elmegreen2005,Elmegreen2007,ForsterSchreiber2009}. Of particular interest is the work of \citet{vdWel2014}, who studied the shapes of star-forming galaxies from $z= 0-2.5$ and found that the fraction of systems that are elongated increases toward higher redshifts and lower masses. For example, at $z \simeq 1$ ($z \simeq 2$) they found that the population of $M_\star \simeq 10^9$ $M_\odot$ ($10^{10}$ $M_\odot$) galaxies is evenly divided between those that are elongated and those that are disky. 

These conclusions are consistent with 
previous observational work on the shapes of galaxies,  indicating that the fraction of elongated galaxies tends to increase with redshift and lower masses. \cite{Ferguson2004} studied a population of galaxies at $z\sim4$ and found a mean ratio of observed semi-major to semi-minor axes of 0.65, suggesting that the sample was not drawn from a spheroidal population. \citet{Ravindranath2006} found spheroidal galaxies at $z>2.5$ were more elongated than more local populations. \citet{Yuma2011}
studied a sample of $\sim 10^{10}$ M$_\odot$ stellar mass galaxies at $z\sim2$ and determined intrinsic 3D shapes via inference to find a mean 3D medium-to-long axis ratio $B/A = 0.61$ and mean 3D short-to-long axis ratio $C/A = 0.28$, suggesting that on average they were elongated. \cite{Zhang2019CANDELS} used CANDELS data to show that star-forming galaxies tend to be elongated at low stellar mass and high redshift and disky at high stellar mass and low redshift, in qualitative agreement with \citet{vdWel2014}.

More recently, \citet{Pandya2024} used JWST data to study the shapes of star-forming galaxies from $z=0.5-8$, and found that many galaxies at $z>1$ with masses consistent with them being progenitors of $L_*$ galaxies today are elongated in shape. For example, they find that the fraction of $M_\star = 10^{9-9.5}\,M_\odot$ galaxies that are prolate rises from $\sim 30\%$ at $z = 0.5$--$1$ to $\sim 50\%$ at $z = 1$--$1.5$, reaching a high fraction of $\sim 70$--$80\%$ from $z = 1.5$ out to $z = 8$. Even among fairly massive galaxies with $M_\star = 10^{10-10.5}$ $M_\odot$ at $z=3-8$, the fraction of systems that are elongated is $\sim 40 - 50\%$.

Such a high fraction of thin, elongated galaxies may point to a problem with standard cosmology.  For example, \citet{Pozo24} used both standard CDM and Wave/Fuzzy dark matter simulations of galaxy formation to show that the elongated fractions observed by JWST disfavor CDM. It is possible that this tension is related to other high-redshift observational challenges, including the possible over-abundance of high stellar mass galaxies  at $z>7$ \citep[e.g.,][]{BK23}, the high fraction of disk/spiral galaxies at early times \citep[e.g.,][]{Kuhn24}, or an elevated UV luminosity density compared with 
modeling predictions \citep[e.g.,][]{Donnan23}. At the same time, these issues could be related to  astrophysical assumptions about dust,  UV light per unit mass, and changing star formation physics \citep[e.g.][]{Finkelstein23,Feldmann25}.

Similarly, it is important to re-emphasize that all estimates of 3D galaxy shapes are subject to inferences based on 2D observables.  \citet{Pandya2024} discuss these issues explicitly. In their work they use a Bayesian model based on a ``toy library" of idealized ellipsoids to back out ellipsoidal shapes from 2D shapes and sizes.  Importantly, they assume that their ellipsoids are uniform and transparent.  As they point out themselves, this is not expected to be the case for real galaxies since dust attenuation should depend on viewing angle and because high-redshift systems typically do not
have uniformly smooth light distributions as is the case in their library of idealized ellipsoids. It is well known that dust affects observables of face-on and edge-on disks differently \citep[e.g.][]{Devoir16}. As an example, using mock FIRE-2 simulations \citet{Klein26} showed that observed axis ratios of edge-on disks depend sensitively on dust fraction (with low dust fractions giving thinner disks at fixed 3D stellar shape). Indeed \citet{Zhang20223DSimulations} directly demonstrate that inclination/dust coupling can mimic structural trends.

In related work, \citet{Vega-Ferrero2024} used a machine learning method trained on TNG50 simulations to classify galaxies observed with JWST and found that a substantial fraction of galaxies that could be naively classified as disks may be intrinsically elongated. Prior to this,  \citet{Ceverino15} and \citet{Tomassetti2016} showed that $\sim 10^9$ $M_\odot$ galaxies at $z \sim 2$ in their simulations are often prolate and live in dark matter halos that are themselves elongated in the same direction as the stellar distribution.  This mass range is roughly what is expected for the mass of the main progenitor of Milky Way size systems at $z=0$ \citep[e.g.][]{SR22}. 

These results raise a question: Did most Milky Way-size galaxies go through a phase of evolution when their stars were forming in elongated shapes?   In this paper, we look specifically at the shape evolution history of Milky Way–like galaxies using a set of 13 zoom-in FIRE-2 \citep{Hopkins2018FIRE-2Formation} simulations. We characterize galaxy shapes using triaxial ellipsoids and measure the axis ratios for all stars, young stars, luminosity-weighted stars, and dark matter at multiple epochs. We also measure the shapes of mono-age populations of stars at $z=0$ and relate them to the shapes that those stars had at the time of their birth. We then compare our results to observational shape distributions and examine the physical properties associated with galaxy elongation, including stellar kinematics, gas inflow, and dark matter structure.

In Section \ref{sec:methods} we discuss our simulations and how the shapes are measured. In Section \ref{sec:results} we present our main results on the 3D shape evolution of simulated Milky Way progenitors. In Section \ref{sec:observations} we produce mock images of our simulated galaxies and construct predicted 2D shape distributions for comparison with observations. In Section \ref{sec:elongation} we examine the physical properties and correlations of galaxy elongation. In Section \ref{sec:conclusions} we summarize our findings and discuss their implications for understanding the early Universe.

\section{Methods}
\label{sec:methods}

\subsection{Simulations}
\label{sec:simulation}

Our simulations were performed as part of the Feedback in Realistic Environments (FIRE) project\footnote{\url{http://fire.northwestern.edu}} and specifically use the FIRE-2 set of feedback prescriptions \citep{Hopkins2018FIRE-2Formation}. In short, the simulations use the code {\texttt{GIZMO}}\footnote{\url{http://www.tapir.caltech.edu/~phopkins/Site/GIZMO.html}}, with hydrodynamics solved using the mesh-free Lagrangian Godunov ``MFM'' method. 
The runs include cooling and heating from a uniform, evolving ionizing background \citep{2009ApJ...703.1416F} and local stellar sources from $T\sim10-10^{10}\,$K; star formation occurs in locally self-gravitating, dense, self-shielding, molecular, Jeans unstable gas. Simulations also include feedback from OB \& AGB mass-loss, SNe Ia \& II, and multi-wavelength photo-heating and radiation pressure with inputs taken directly from stellar evolution models.

In this paper, we analyze the 3D geometry of 13 simulated galaxies, all of which reach a mass similar to that of the Milky Way at $z=0$. Seven of these are part of the Latte suite \citep{wetzel2016reconciling,garrison-kimmel2017not-so-lumpy,hopkins2017anisotropic,garrison-kimmel2019the-local} and six are part of the ``ELVIS on FIRE" project, which were simulated in pairs similar to the Milky Way and M31  \citep{garrison-kimmel2019the-local,garrison-kimmel2019star}.  Other than the ELVIS runs being associated with Local-Group like pairs, the main difference is that the Latte suite is run with initial baryonic particle masses $\sim 7000$ $M_\odot$ that are about twice as large as those in the ELVIS suite, $\sim 3500$ $M_\odot$. There were no pre-conditions on the Latte suite zooms other than $z=0$ mass selection.  Given that they were chosen to live in Local-group type environments, the ELVIS runs are likely slightly over-dense compared to the average system of that mass. More details of these simulations can be found in the associated references above.

\subsection{Measuring and Characterizing Simulated Galaxy Shapes}
\label{subsec:methods}

Axis ratios of stars in simulated galaxies have been calculated with a variety of methods including 3D isodensity surfaces, mass eigentensors, or minimum-volume-enclosing ellipsoids \citep[e.g.][]{Ceverino15,Tomassetti2016,Thob2019, Zhang20223DSimulations,Valenzuela2024,Keith2025}. In this work, we adopt a shape tensor approach \citep[e.g.][]{2005ApJ...627..647B,Allgood2006} and specifically use a reduced mass eigentensor \citep{2019MNRAS.484..476C,Klein26}:
\begin{equation}
S_{ij} = \frac{\sum_n r_{\text{ell},n}^{-2} m_n x_{n,i} x_{n,j}}{\sum_n m_n},
\label{eq:R}   
\end{equation}
with the indices $i$ and $j$ each running over the three spatial dimensions and the sum covering $n=1, 2, ... N$ star particles within the volume of choice (see below). The quantity $m_n$ represents the star particle's mass, and $x_{n, i}$, $x_{n, j}$ are pairs of coordinate distances in the $i$ and $j$ directions. Diagonalizing the tensor yields eigenvalues $\lambda_1 \geq \lambda_2 \geq \lambda_3$, from which the axis lengths may be associated as $A, B, C \propto \sqrt{\lambda_1}, \sqrt{\lambda_2}, \sqrt{\lambda_3}$. Important in our analysis is the quantity $r_{\text{ell},n}^{-2}$, which allows us to weigh each star particle inversely with its elliptical distance from the galaxy center:
\begin{equation}
    r_{\text{ell}} = \sqrt{x^2 + \frac{y^2}{(B/A)^2} + \frac{z^2}{(C/A)^2}}.
\end{equation}
Here, $A, B, C$ represent the axis lengths of the current iteration. Without this weighting, particles with large radii within the volume tend to drive inferred shapes to be more spherical, limiting the dynamic range of the associated axis ratios. This aligns with work by \citet{Valenzuela2024}, which finds that this weighting provides a strong measurement for the shape of a galaxy within a boundary. Independently, \citet{ForouharMoreno2026} demonstrate using the COLIBRE simulations that iterative inertia tensors recover significantly larger fractions of flattened and disc-like galaxies than non-iterative approaches, regardless of aperture size, further validating our methodology.

We calculate galaxy shapes iteratively by first considering all star particles within a spherical volume of radius equal to one tenth of the host halo virial radius\footnote{We define the virial mass and radius of our galaxies following the definition in \citet{Bryan1998}. Note that the virial radius evolves with time for any particular galaxy.}, $A=B=C=0.1\,r_{\mathrm{vir}}$.
The reduced mass eigentensor $S_{ij}$ gives us an initial set of eigenvectors $\vec{A}, \vec{B}, \vec{C}$ and eigenvalues $\lambda_1, \lambda_2, \lambda_3$. The rotation matrix $R^{-1}$ is used to rotate the galaxy from the simulation axes to the new eigenbasis, where $R = [\vec{A},\vec{B},\vec{C}]$. In the first iteration, the square roots of the eigenvalues are scaled so that the largest value matches the distance from the center to the farthest star particle in the volume. This produces a new set of principal axes $\vec{A}^\prime, \vec{B}^\prime, \vec{C}^\prime$. After the first iteration, we scale the lengths of the axes by a factor of 
$({A'B'C'}/{ABC})^{1/3}$ to preserve volume. This process is repeated until the axis ratios $B/A$ and $C/A$ converge to within $10^{-4}$ of the previous iteration. To precisely define the final 3D ellipsoidal shape, extraneous particles are excluded by scaling the ellipsoid down until it retains 90\% of its original mass. This adjustment minimally affects the shape due to the factor $r_{\text{ell}}^2$ in the denominator. We define the stellar mass of each galaxy as the total stellar mass within this ellipsoidal volume.

\begin{figure}
    \centering
    \includegraphics[width=\columnwidth]{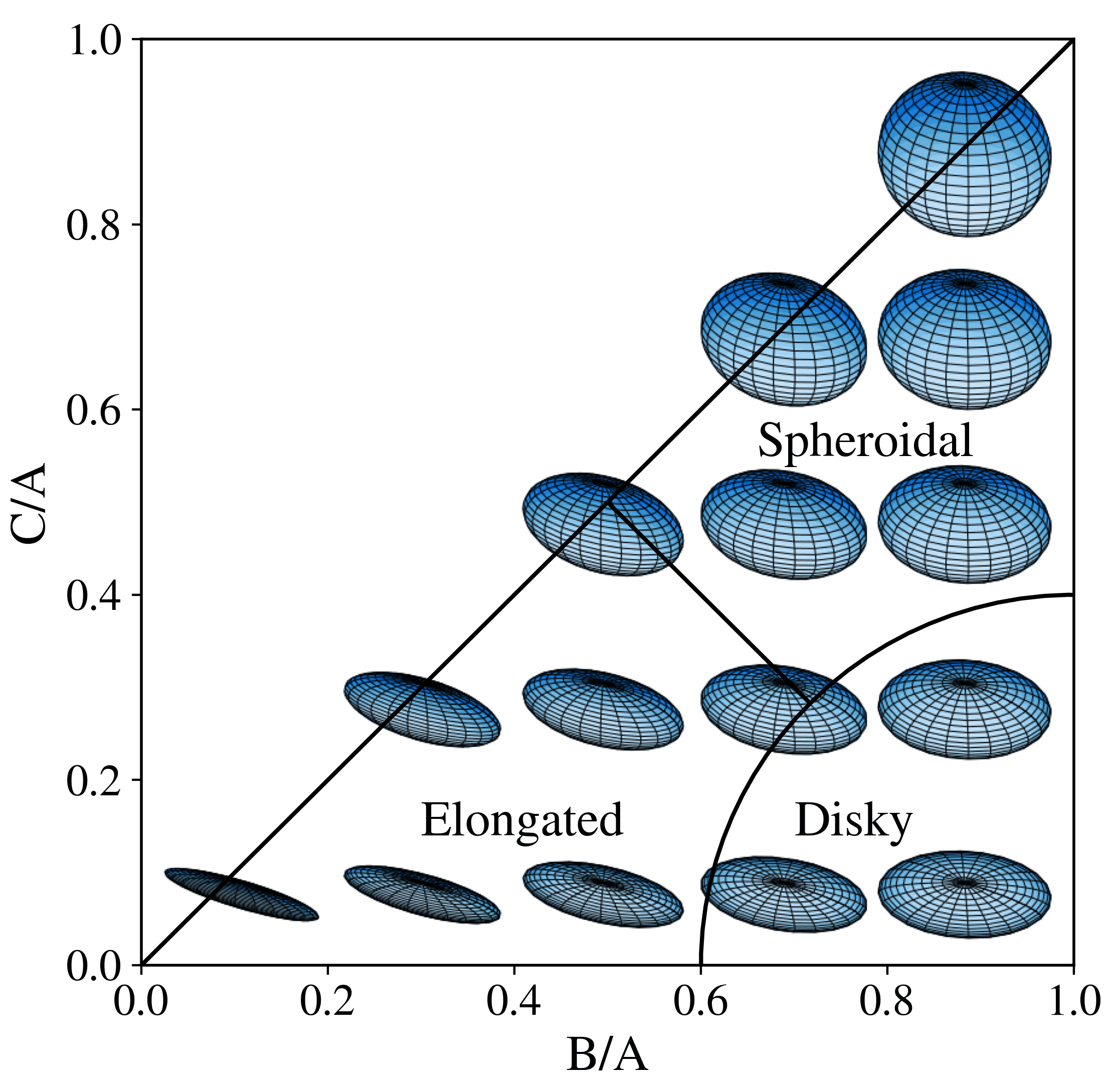}
    \caption{A sample of ellipsoids that demonstrate the meaning of our $C/A$ and $B/A$ parameter space, where $C < B < A$. The defined regions for disky, spheroidal, and elongated shapes are indicated in the figure. }
    \label{fig:baca_mapping}    
\end{figure}

Following \citet{Zhang2019CANDELS}, we categorize galaxies into disky, spheroidal, and elongated shapes based on their relative axis ratios $C/A$ and $B/A$. Figure \ref{fig:baca_mapping} illustrates how we separate galaxies into each of these categories. Idealized ellipsoids are depicted as examples for comparison. 

\subsection{Shape Selection and Weighting} 

We track shapes using five distinct weighting schemes and choices of components. The first four explore the shape of the main progenitor of each galaxy through cosmic time using 18 snapshot outputs. The snapshots are spaced approximately 750~Myr apart from $z=7$ to $z=0$. Throughout the paper, lookback time refers to the time elapsed between a given snapshot and the present day; for example, $z=7$ corresponds to a lookback time of 13.0~Gyr.

In the first case, we measure the mass-weighted shape of all stars.  This gives an integrated view of the galaxy's structure at each epoch. Second, we measure shapes of young stars (ages $< 500$~Myr). This provides the shape of newly-made stars and is independent of how the older stars in the galaxy are arranged at any given time.  This allows us to connect the shape distribution of stars soon after they were made to how the shape of those same stars evolves over time (see below).\footnote{We have looked at stars younger than 100~Myr and find similar results, but the 500~Myr age range provides consistently enough star particles ($\geq 1000$) to measure shapes accurately at all lookback times considered, as discussed in Section~\ref{subsec:measure}.}  Third, we measure luminosity-weighted shapes by replacing the particle mass $m_n$ in Equation~\ref{eq:R} with its $g$-band luminosity $L_n$; this is quasi-observable because it tracks light, although real observations are affected by dust and other projection factors we address in Section~\ref{sec:observations}. Luminosity-weighting systematically biases shape measurements relative to mass-weighting because younger stars are kinematically distinct from the older stellar population \citep{ForouharMoreno2026}. Fourth, we apply the same algorithm to dark matter particles rather than stars, within both our standard volume ($0.1\,r_{\rm vir}$) and an extended radius of $r_{\rm vir}$.

Finally, our fifth measure is ``archaeological.'' In this case we measure the shapes of 18 mono-age stellar populations at $z=0$.  Each age population is chosen to match the lookback times of the snapshots binned in 500~Myr age intervals. Intuitively, the archaeological measurement captures the shapes of where stars are now, while the young-stars measurement can be thought of as the shape of those same stars when they were born.

\subsection{Measurement Accuracy}
\label{subsec:measure}
Measuring the shapes of our galaxies accurately at $z=0$ is fairly straightforward because each simulation contains millions of star particles. However, as we examine subsets of particles in progenitor galaxies at earlier times, the particle count declines rapidly. We perform the following experiment to assess the minimum number of particles required to accurately measure the shape of a system. We first measured the shapes of several highly-resolved galaxies at $z=0$ including galaxies that are spheroidal, elongated, disky, and undergoing mergers. Specifically we measured shapes within $0.1\,r_{\mathrm{vir}}$ and recorded this as their ``true'' shape. We then progressively removed fractions of particles and analyzed the deviations in the resulting shape measurements. For each fraction, we removed a random subset of particles and recalculated the shape 100 times to quantify the deviation. We increased the fraction of star particles removed until the standard deviation in derived axis ratios, divided by the ``true'' shape, exceeded 0.05. Our findings indicate that this threshold was consistently exceeded when fewer than 1,000 particles remained. Thus, we adopt a lower limit of 1,000 particles as the minimum necessary to reliably measure the galaxy's shape.\footnote{In practice, the need for 1,000 particles is what motivated us to choose a 500 Myr cutoff for stars we classify as ``young''. A younger age limit would have resulted in fewer stars at some of our higher redshift timesteps.} 

\begin{figure}
    \centering
    \includegraphics[width=0.5\textwidth]{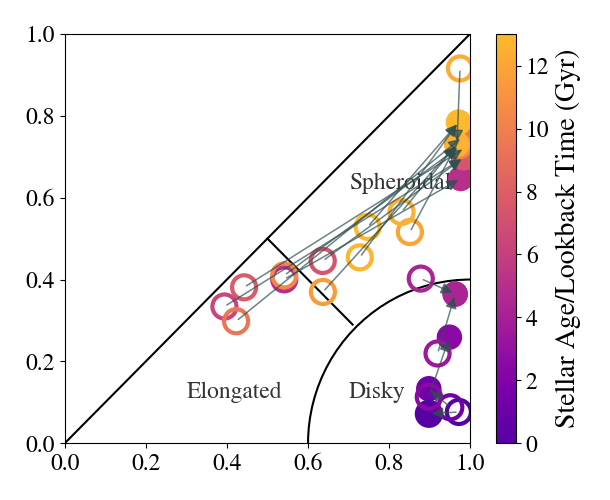}
    \caption{Shapes of Thelma's stellar populations at formation (open) and at $z=0$ (filled), binned by age, in the parameter space introduced in Figure \ref{fig:baca_mapping}. An arrow connects each stellar population shape at formation to its shape at the present day. The color bar maps to stellar age (or, equivalently, lookback time to their formation). At early times, young stars often resided in elongated or triaxial-spheroidal configurations when they formed. The shapes of the same populations today are much ``rounder" and more symmetric about their short axes, with $B/A \sim 0.9$. }
    \label{fig:Thelma_pastpresent}
\end{figure}

\begin{figure*}
    \centering
    \includegraphics[width=\textwidth]{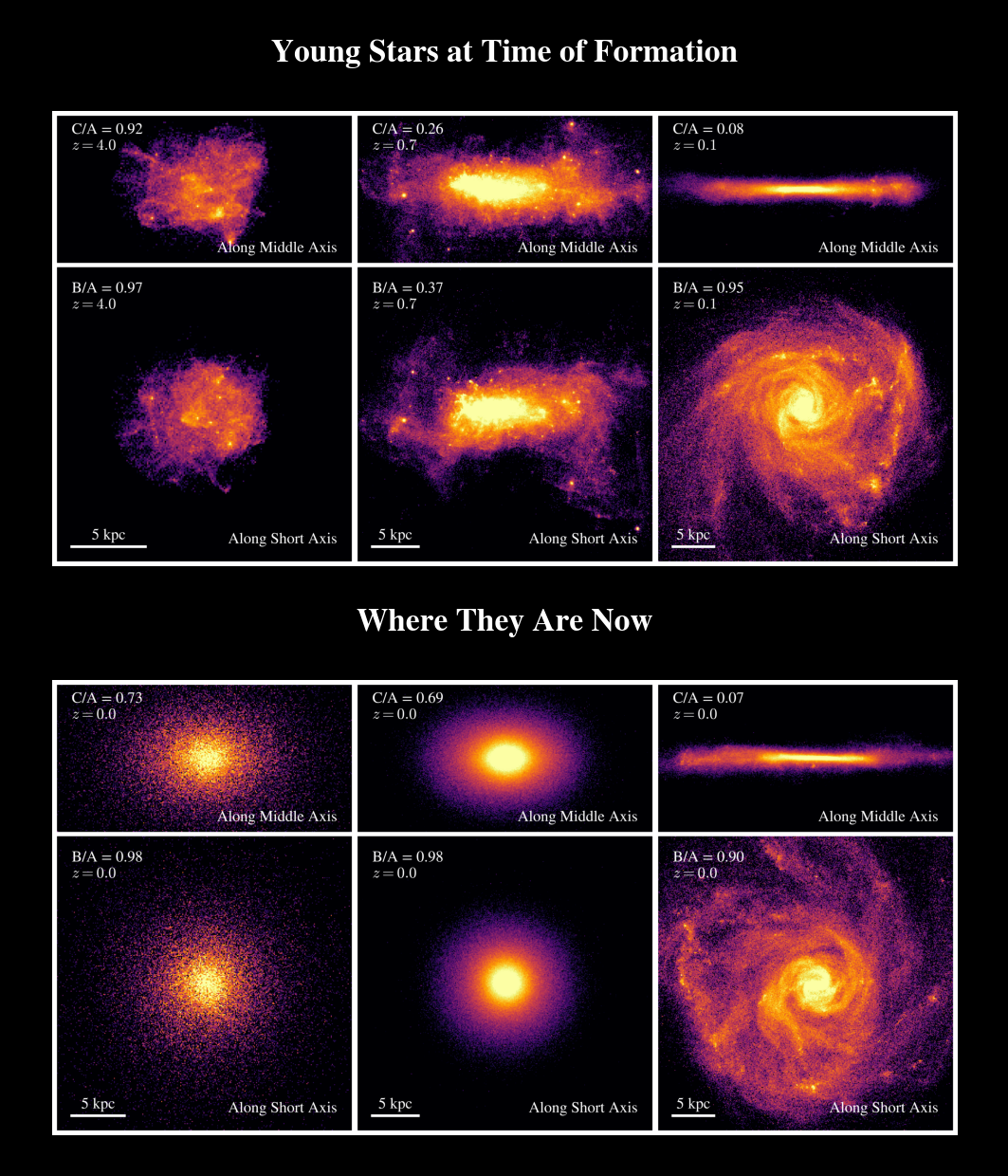}
    \caption{Density projections of stellar populations younger than 500 Myr near their times of formation (top set) and at $z=0$ (bottom set). In the top set of images, we show middle- and short-axis projections of young stars at $z=4.0$ (left, 12.2 Gyr lookback), $z=0.7$ (middle, 6.4 Gyr lookback), and $z=0.1$ (right, 1.3 Gyr lookback). The bottom set of images shows the same stars, now at $z=0$, along the same respective axis orientations (though the axes are calculated separately at $z=0$). We see that the initially spheroidal population (left) evolves into a spheroidal shape at $z=0$. The elongated stellar population (middle) at $z=0.7$ has evolved into a flattened spheroid over the past 6.4 Gyr. The stars that formed in a thin disk at $z=0.1$, or 1.30 Gyr ago (right), still remain in a thin disk today. Regardless of their initial shapes, in all cases the $z=0$ populations are quite symmetric (circular) in short-axis projections. }
    \label{fig:thelmaevolution}
\end{figure*}
\section{Results}
\label{sec:results}

In what follows, we first present shape evolution results for a single simulated galaxy, Thelma, and in the following subsection we discuss sample-wide results. We begin with a single galaxy for the sake of clarity. Thelma is fairly typical of the galaxies in our suite in that it forms a prominent disk at $z=0$. We present Thelma as a representative case; similar results for the remaining 12 galaxies are shown in Appendix~\ref{sec:appendix2}.

\begin{figure*}
    \centering
\includegraphics[width=\textwidth]{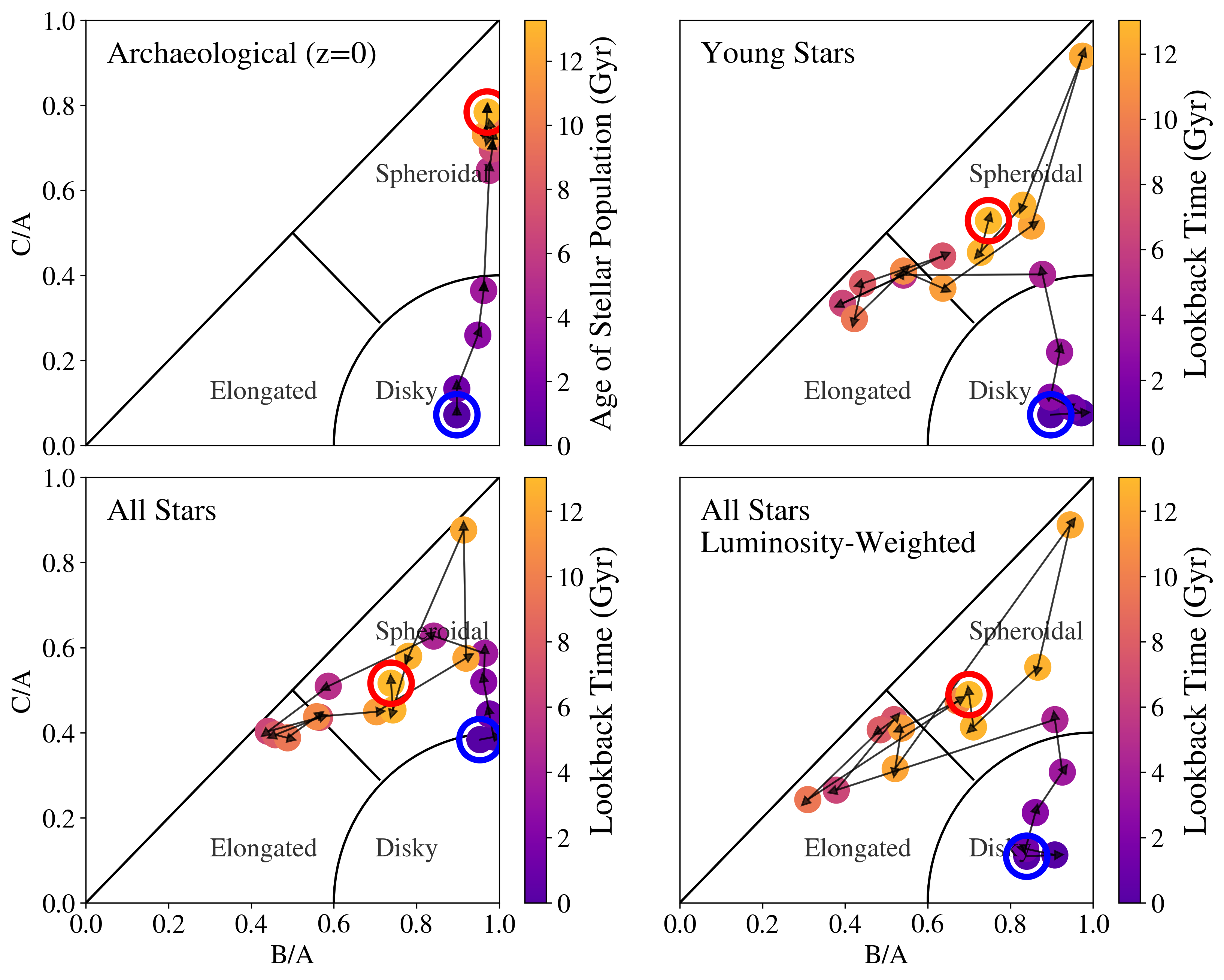}
    \caption{
    Shape evolution for one example Milky Way–like galaxy, Thelma. Each panel presents axis ratios as defined in \S \ref{subsec:methods} broken into the shape categories defined in Figure \ref{fig:baca_mapping}. The top left panel shows the shapes of mono-age stellar populations at $z=0$. The color of each filled circle maps to stellar age as defined by the color bar. Beginning with the oldest population (red circle), arrows point to the next oldest age bin, up to the youngest age bin circled in blue. The top right shows the configurations of the same stellar populations at the time they were born: the axis ratios of young stars at various lookback times. The bottom left panel shows the same measurement, but now for all stars in the main progenitor at various lookback times. In the bottom right we show the shape of the main progenitor, now weighted by $g$-band luminosity.
}
       \label{fig:Thelmapoints}
\end{figure*}

\subsection{Case Study: Thelma}

Figure \ref{fig:Thelma_pastpresent} presents the evolution of the shapes of individual stellar populations in Thelma. The shapes are tracked in the shape parameter space ($C/A$ vs. $B/A$) introduced in Figure \ref{fig:baca_mapping}.

In Figure~\ref{fig:Thelma_pastpresent}, open circles show the birth-time shapes of young stellar populations, connected by arrows to the shapes of the same populations at $z=0$ (filled circles), with color encoding lookback time.

Young stars at late times ($< 4$ Gyr lookback) tend to inhabit disky configurations and remain so at $z=0$. However, young stars are often arranged in elongated or triaxial/spheroidal configurations at lookback times greater than $\sim 4$ Gyr. In almost every case, the same stellar populations wind up drastically more symmetric about their minor axis at $z=0$ ($B/A \simeq 0.9$), and none evolve into elongated configurations by the present day. Also of interest is that at intermediate lookback times ($\sim 4-6$ Gyr) some of the initially elongated configurations evolve into axisymmetric disks, while most of the elongated configurations (which occurred slightly earlier, $\sim 6-8.5$ Gyr ago) evolve into axisymmetric spheroidal shapes.

An example visualization of this kind of evolution in shape is shown in Figure \ref{fig:thelmaevolution}. The top panels show 2D density maps of young stars near the time of their birth and the bottom panels show the same populations at $z=0$.  Specifically, the top set of panels shows the middle axis (upper) and short axis (lower) orientations of young stars in Thelma at $z=4$ (left), $z=0.7$ (middle) and $z=0.1$ (right). The lookback times correspond to 12.2 Gyr, 6.4 Gyr, and 1.3 Gyr, respectively.  The bottom panels show the same stars at $z=0$, oriented along their $z=0$ middle and short axes.  Of particular note is that, while the young stars at $z=0.7$ were arranged as an elongated ``pickle", those stars evolve into a flattened spheroidal shape at $z=0$. In contrast, the population of young stars at $z=4.0$ is spheroidal in shape both at formation and at $z=0$.  Similarly, young stars at $z=0.1$ form in the shape of a disk and remain in a disk at $z=0$. 

\begin{figure*}
\centering
\includegraphics[width=0.85\textwidth]{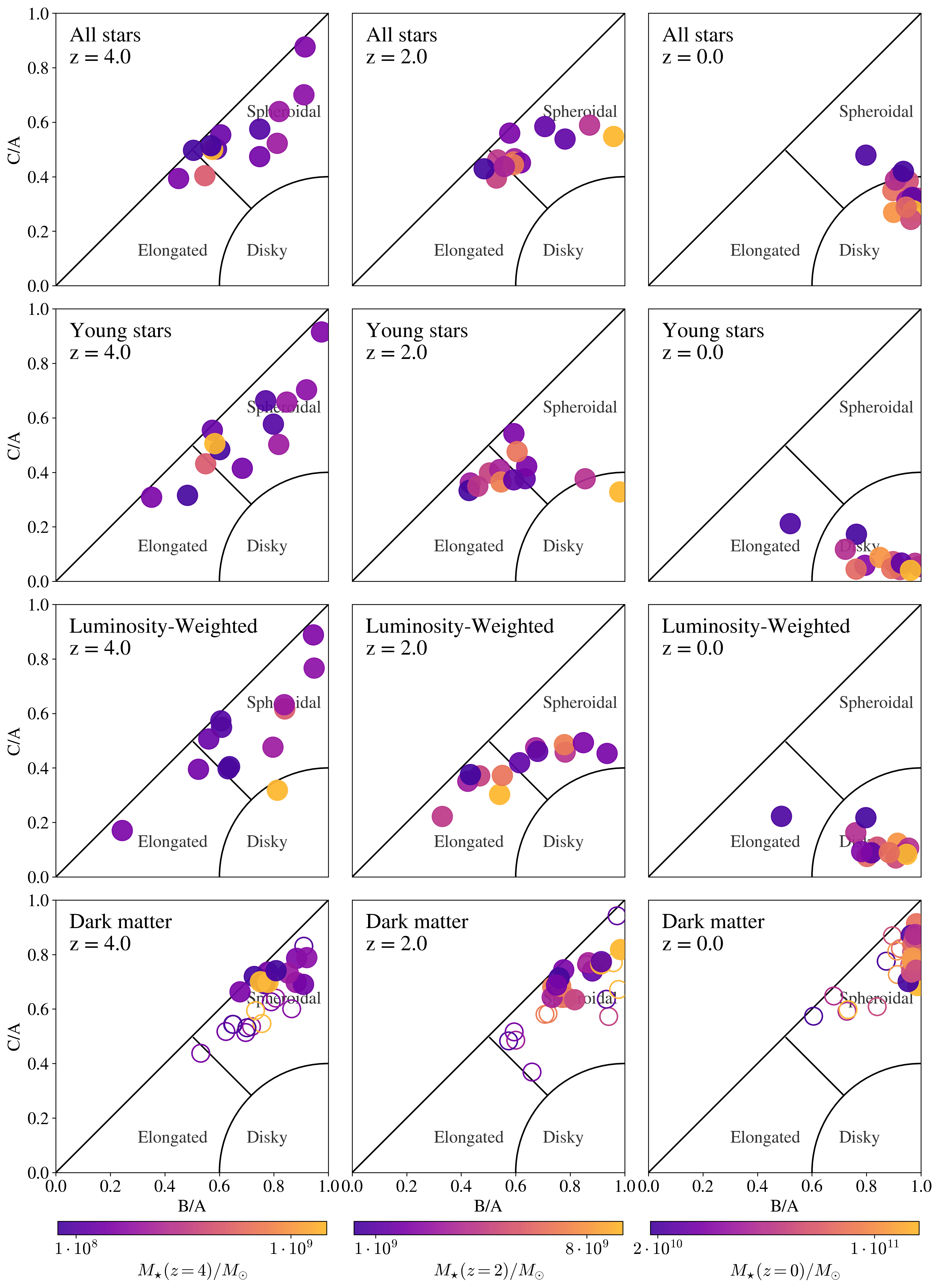} 
\caption{
The 3D shape evolution of the main progenitors of all of our FIRE-2 galaxies with Milky Way masses at $z=0$. Each panel uses the classification scheme introduced in Figure \ref{fig:baca_mapping}. The three columns show progenitor shapes at three redshifts: $z = 4.0, 2.0,$ and $0.0$, from left to right. The rows show the shape measurements for different components: all stars (top),  young stars (second), luminosity-weighted stars (third), and dark matter (bottom). The open circles in the bottom panel show dark matter shapes measured out to $r_{\mathrm{vir}}$ rather than our standard $0.1r_{\mathrm{vir}}$. The color bar maps to the logarithm of the stellar mass of the all stars sample so that the same galaxy has the same color in a given column. Qualitatively, galaxy shapes tend to be more elongated or spheroidal at early times, and evolve into disky configurations at $z=0$. The dark matter is almost always spheroidal with the exception of one point at $z=4$. 
}
\label{fig:allgalaxies}
\end{figure*}

\begin{figure*}
    \centering
    \includegraphics[width=\textwidth]{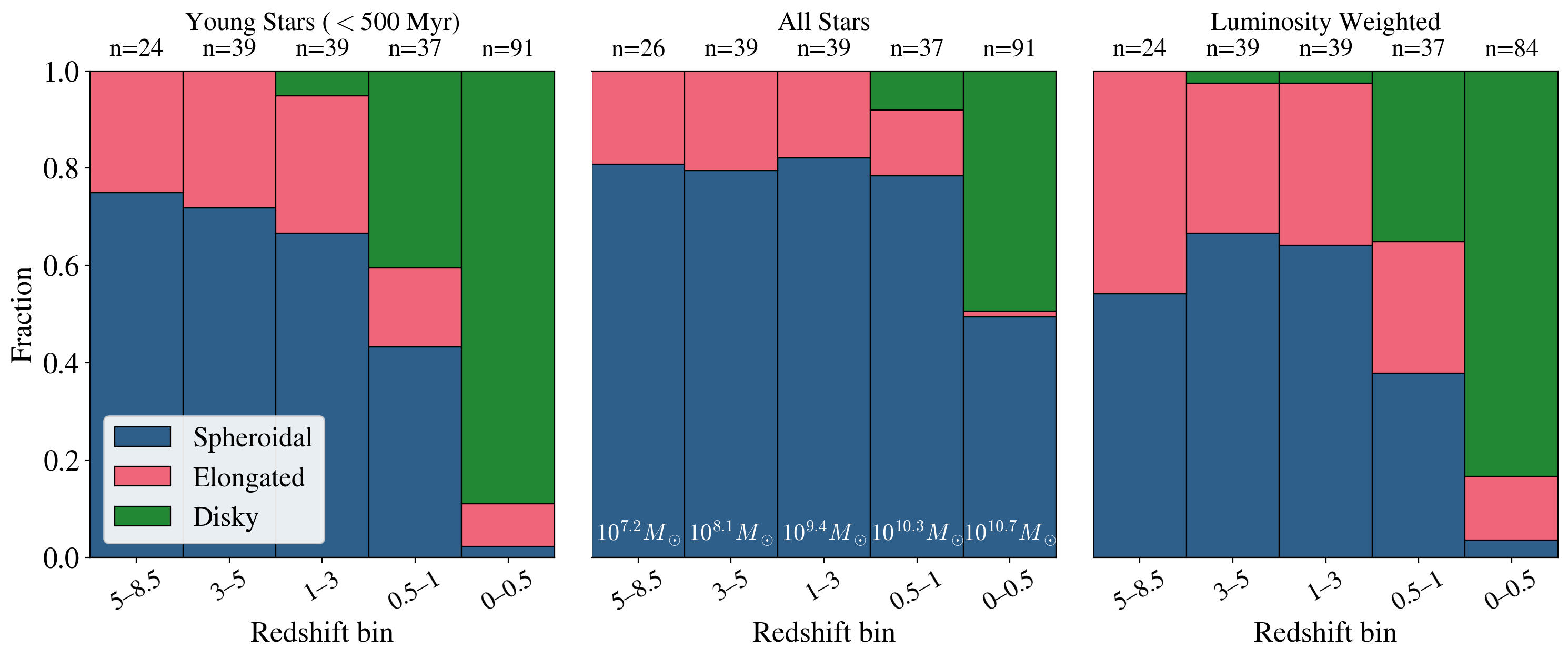}
    \caption{The fraction of snapshots classified as spheroidal (blue), elongated (pink), and disky (green) as a function of redshift, shown separately for young stars ($<500\,\mathrm{Myr}$, left), all stars (middle), and luminosity-weighted stars (right). Sample sizes $n$ are indicated at the top of each bin; for the all-stars panel, the mean stellar mass of each bin is shown on the bottom. At high redshift, young star populations are a mix of spheroidal and elongated configurations, with the elongated fraction peaking near $z \sim 1$--3. Disky configurations are essentially absent at $z \gtrsim 1$ and come to dominate only at $z \lesssim 0.5$. The all-stars panel evolves more gradually because it integrates over all stellar populations, including stars born before the disk phase. The luminosity-weighted panel broadly tracks the young-star panel, with a more prominent elongated fraction at intermediate redshifts.}
    \label{fig:shape_fractions}
\end{figure*}

Figure \ref{fig:Thelmapoints} presents a related examination of Thelma's shape evolution,  again in the $C/A$ vs $B/A$ parameter space.  The upper left panel reproduces the shapes of $z=0$ stellar populations binned by age, shown in Figure \ref{fig:Thelma_pastpresent}. We see that the oldest populations ($\gtrsim 8$ Gyr) reside in spheroids ($C/A \sim 0.6$). The youngest stars ($\lesssim 2$ Gyr) inhabit thin disks ($C/A \sim 0.1$). Intermediate-age stars ($\sim 4$ Gyr) reside in thicker disks ($C/A \sim 0.3$). Notably, stars of all ages are very symmetric about the short axis, with $B/A \sim 0.9-1$.

The top right panel of Figure~\ref{fig:Thelmapoints} shows the shapes of the same stellar populations {\em at the time they were born}, now connected as a time sequence. This allows us to more directly track how the shapes are changing in time. Young stars at lookback times $\lesssim 4$ Gyr tend to inhabit disks. However, at intermediate lookback times and later, young stars are arranged in either spheroidal or elongated configurations. Young star shapes in Thelma tend to move within the spheroidal region for the first $\sim 5$ Gyr of cosmic time.  They subsequently have elongated configurations for the next $\sim 4$ Gyr. During these early times, axis ratios deviate quite far from axisymmetry, with $B/A \sim 0.4 - 0.8$. This is striking in comparison to the shapes that those same stars inhabit at $z=0$ (top left). Again, even though these stars are often formed in elongated or triaxial arrangements, they relax into quite round, axisymmetric configurations by the present day. 

The bottom left panel of Figure \ref{fig:Thelmapoints} shows the shape of Thelma's main progenitor over time, measured using the mass-weighted eigentensor of all stars. At $z=0$, the mass-weighted shape is characterized as a thick disk ($C/A \sim 0.4$) and can be interpreted as a weighted average of the archaeological age bins shown in the top left panel. Compared to the young stars' shapes, there is much less variation in overall shape of the total star population with time, as might be expected due to the averaging over a larger number of particles. Nevertheless, even the all-star shape goes through periods of elongation.  As with young stars, the main progenitor is much less axisymmetric at intermediate times than we see in intermediate-age populations at $z=0$.

The bottom right panel presents a quasi-observable measure of shape evolution: the $g$-band luminosity weighted axis ratios at the same lookback times. The luminosity-shape evolution is qualitatively similar to the young stars' shape evolution: the shapes move from mostly spheroidal at the earliest times to elongated at intermediate times.  At late times, the luminosity-weighted shapes become disky.  Interestingly, the luminosity-weighted points show some instances of extreme elongation, including one case with $B/A \sim C/A \sim 0.3$.

Before moving on to present a summary of sample-wide trends, we note that in Appendix \ref{sec:appendix2} we present similar plots to those shown in Figures \ref{fig:Thelma_pastpresent} and \ref{fig:thelmaevolution} for all of our other simulations (Figures \ref{allgalaxiespastpresent} - \ref{allm12AllStars}).  For all but one of those twelve, the results are similar to those of Thelma. The lone exception is m12z, which is experiencing a late-time merger. For the rest, which we refer to as ``Milky Way analogs," young stars form in disks at late times, and at early times, their luminosity-weighted and mass-weighted shapes oscillated between elongated and spheroidal over $\sim 1-3$ Gyr timescales. The outlier, m12z, has an elongated shape at $z=0$, driven by a merger. This may be a clue that mergers play a role in the elongated phases prevalent among our sample at higher redshifts; we return to this question in Section~\ref{sec:dm}.

\subsection{All Galaxies}
Figure \ref{fig:allgalaxies} presents a summary of the 3D shape evolution of our 13 FIRE-2 galaxies sampled at three example redshifts. The columns present shape results for our Milky Way–size galaxies ($z=0$, right column) and their main progenitors at redshifts $z=2$ (middle column) and $z=4$ (left column).  The color of each point in all rows maps to the total stellar mass of the galaxy at the associated redshift, as encoded by the color bars on the bottom of each row. This means that for each column, the same galaxy has the same color in all four rows.

The top row of Figure \ref{fig:allgalaxies} presents shapes determined using all stars in the galaxy at the given redshift. At $z=0$ (upper right), all but one galaxy has an axisymmetric ($B/A \sim 0.95$) disky ($C/A \lesssim 0.4$) configuration.  Note that when the shapes are measured using all stellar mass like this, even the $z=0$ disks are rather thick ($C/A \simeq 0.3$). The one  case that is not a disk is m12z, which is one of the least massive (purple) points.  This galaxy is undergoing a merger at $z=0$. At $z=2$ and $4$, the mass-weighted shapes are typically flattened spheroids ($C/A \sim 0.5-0.9$, $B/A \sim 0.6-0.8$), although we also see a few progenitor galaxies with elongated shapes, with $C/A \sim B/A \sim 0.4$.

The shape evolution for young stars  (second row in Figure \ref{fig:allgalaxies}) is much more dramatic. At $z=0$, young stars tend to reside in very thin disks, with $C/A \sim 0.1$, in all but one of our galaxies. As discussed above, the one exception is the lower-mass (purple) galaxy m12z, which is undergoing a merger, and has an elongated shape in young stars.  Compared to the mass-weighted shapes at $z=0$,  the young stars distribution is less symmetric about the short axis ($B/A \sim 0.7-0.9$) than the all-stars distributions ($B/A \sim 0.9-1$, top panel), even though the young stars reside in much more flattened disks ($C/A \sim 0.1$).

The young stars at $z=2$ (second row, middle panel of Figure \ref{fig:allgalaxies}) have a dramatically different arrangement from the young stars at $z=0$: most are elongated and none are in the form of thin disks. At $z=4$ (left panel), young stars show a variety of shapes, from fairly spheroidal ($C/A \sim B/A \sim 1$) to very elongated ($C/A \sim B/A \sim 0.3$). Although there is variance, we do see more spheroidal configurations among the young stars at $z=4$ than we see at $z=2$. 

The third row of Figure \ref{fig:allgalaxies} presents our ``quasi-observable" shape analysis, weighted by $g$-band luminosity. 
Qualitatively, these shapes are more similar to the young stars measurements than the all-stars configurations, but we see more variance. In particular, at $z=2$, we see that several of the $g$-band shapes are fairly elongated. One progenitor of note at $z=4$ is extremely prolate, with $C/A \sim B/A \sim 0.2$ and we even have one $g$-band progenitor at $z=4$ shaped like a thick disk. 

The last row of Figure \ref{fig:allgalaxies} focuses on the shapes of the galaxies' dark matter component.  The solid points show shapes measured with our standard volume of $0.1\,r_{\mathrm{vir}}$, and the open points are measured within a larger volume out to $r_{\text{vir}}$, for comparison.  Unlike the stellar components, the inner dark matter shapes are always quite spherical. The larger-scale dark matter shapes (out to $r_{\mathrm{vir}}$, open) show more variance but are always within the spheroidal region, with the exception of one point (m12f at $z=4$). The dark matter in the vicinity of the galaxies (solid points) displays a gradual transformation from triaxial spheroids at high redshift to near-perfect spheroids at lower redshifts, with $B/A \sim 1$. We do not see a systematic correlation between the elongation of dark matter shapes and the elongation of stellar populations, except for the most extreme objects at $z=4$ (light purple, m12f) and $z=0$ (dark purple, m12z), where the virial-radius shapes are also the most extreme. For the case of m12z, we know that this is a merger.

\section{Predictions and Connections to Observations}
\label{sec:observations}

Although our sample is not a fair census of the universe — being 
drawn specifically from Milky Way-mass systems at $z=0$ — we make an effort here to compare the progenitor shape distributions to current estimates from larger (fair) samples and to make associated predictions.  In particular, we focus on comparing our results to those of \citet{Pandya2024}, who have a galaxy sample with stellar masses $10^9-10^{10.5}$ M$_\odot$ over a redshift range $z=0.5-8$.  

\subsection{3D shapes vs. observations}

Because shape distributions change with mass, it is important to emphasize that our predicted sample has a mass distribution that evolves with redshift. Our stellar masses peak at low redshift, spanning $\sim 10^{10}$-$10^{11}\,M_\odot$ at $z=0$--1. From $z = 1 - 2$, our sample spans a stellar mass range $\sim 10^{9} -10^{10.3}$M$_\odot$. Above $z = 3$, all of our progenitors have masses that fall below those studied directly in \citet{Pandya2024}, although we present results for the sake of completeness and prediction.

Figure~\ref{fig:shape_fractions} shows a statistical summary of our predicted shape evolution fractions over several redshift bins (bottom axis), measured using young stars (left), all stars (middle), and luminosity-weighted stars (right). Note that we only have 13 galaxies total, each of which has multiple progenitor redshift snapshots; the $n$ values at the top of each bin in Figure~\ref{fig:shape_fractions} count the total number of galaxy--snapshot pairs included in that bin. Given this sample size, all quoted shape fractions carry binomial $1\sigma$ uncertainties of order $\pm 5$--$10\%$. The masses listed along the bottom of the middle panel in white are the median stellar masses for the sample in those redshift bins. 

In comparing these results to \citet{Pandya2024}, the luminosity-weighted results (right panel) provide the most appropriate basis, although the comparison is not perfect. \citet{Pandya2024} assume that the ``true" galaxy population is a distribution of perfectly smooth and uniformly light-emitting triaxial ellipsoids, and they use forward modeling with this assumption to infer 3D shape fractions from the observed 2D axis ratios and 2D semi-major-axis sizes of real galaxies. 

With these caveats in mind, we compare to \citet{Pandya2024} in three redshift--mass bins where the overlap is genuine. At $z=0.5-1$, our progenitors span $M_\star \sim 10^{10}$--$10^{11}\,M_\odot$, overlapping with the \citet{Pandya2024} $10^{10}$--$10^{10.5}\,M_\odot$ bin. We find an elongated fraction of $\sim 20\%$ in luminosity-weighted shape, consistent with their reported $\sim 25\%$ at the same mass and redshift. At higher redshift, $z=1-3$, our sample masses have shifted down to $\sim 10^8 - 10^{10}$ M$_\odot$ and we measure an elongated fraction of $\sim 35\%$. At $z \sim 2$, \citet{Pandya2024} report a $\sim 40\%$ elongated fraction among their lowest-mass galaxies. Given the caveats discussed above, these are reasonably consistent fractions. We also find that the luminosity-weighted elongated fraction generally rises with redshift, reaching $\sim 45\%$ in the highest redshift bin. This qualitative increase in elongated fraction with redshift is consistent with the results of \citet{Pandya2024}, though the mass range in our sample at this redshift is far below the lower mass limit in their paper.

\subsection{2D shape predictions}
Given the uncertainties associated with extracting 3D shape distributions from 2D data, we present 2D shape distributions directly for our progenitor samples using mock images. Mock images are generated following the methodology of \citet{Klein26}, to which we refer the reader for full details. Briefly, we use the Monte Carlo radiative transfer code \texttt{SKIRT} \citep[][]{Camps2015, Camps2020} to produce mock images in the rest-frame $g$-band, with a fixed dust-to-metal ratio of 0.1. For each galaxy, we generate random viewings and use \texttt{AstroPhot} \citep{2023MNRAS.525.6377S} to fit a S\'ersic surface brightness profile \citep{Sersic1963Influence}:
\begin{equation}
  I(r(x,y)) = I_e \exp \bigg\{ -b_n \bigg[ \Big( \frac{r}{R_e}\Big) ^{1/n}- 1 \bigg] \bigg\}.
\end{equation}
Here, $r(x,y)$ is a rotated elliptical radius defined by an orientation angle and a minor-to-major axis ratio $q \equiv b/a$.

Before showing sample-wide predictions for $q$ distributions, we begin by selecting three example galaxies representing each of our three shape classes. We create 30 mock images for each one along random sight lines and build a $q$ distribution for each galaxy individually. 

Specifically, we use 3D luminosity-weighted shapes to choose 1) a disky example with $C/A = 0.11$ and $B/A = 0.84$ (Thelma at $z=0$); 2) an elongated example with $C/A = 0.26$ and $B/A = 0.38$ (Thelma at $z=0.7$); and 3) a spheroidal example with $C/A = 0.95$ and $B/A = 0.89$ (Thelma at $z=4$).  

\begin{figure}
    \centering
    \includegraphics[width=\columnwidth]{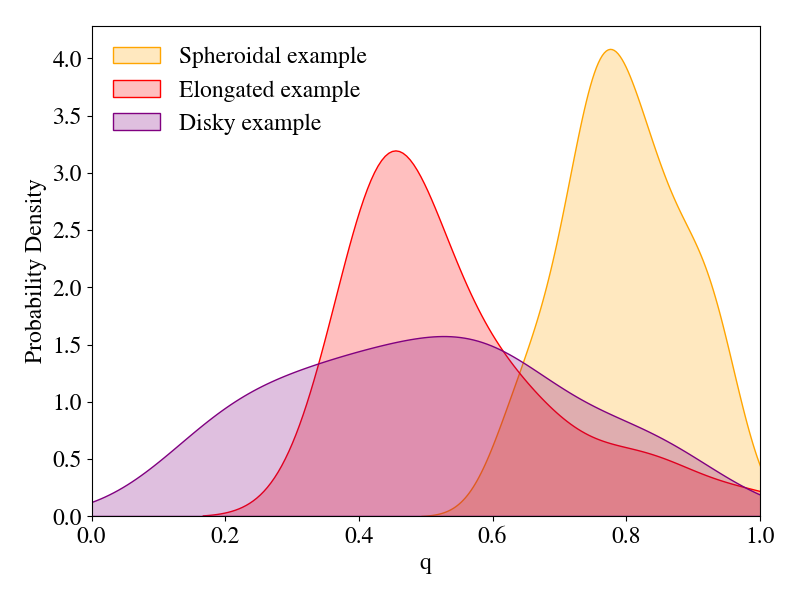}
    \caption{Observed 2D shape distributions for a  spheroidal, elongated, and disky galaxy. We choose one galaxy of each class and project it from 30 random viewing angles. The horizontal axis plots the $q \equiv b/a$ parameter obtained from the S\'ersic fits to our mock images, while the vertical axis plots the probability density. The solid outlines are the KDE plots generated from the $q$ distribution for each class. The simulated spheroidal case produces a narrow distribution peaked at round shapes, with $q \sim 0.8$. The elongated case peaks at $\sim 0.45$, with a tail towards larger $q$. The disky example produces the full range of $q$ values.}
    \label{fig:pure_shape_distribution}
\end{figure}

Figure \ref{fig:pure_shape_distribution} shows the results of this idealized exercise. The Spheroid example produces a $q$ distribution highly peaked towards circular shapes, with almost all projections yielding $q > 0.6$. This is expected for a nearly spherical object. The disky example, on the other hand, has a broad distribution, with the extremes set by it being viewed edge-on ($q \sim 0.1$) or face-on ($q \sim 0.9$). The elongated example produces a skewed distribution, peaking at $q \sim 0.45$, with a tail towards $q \sim 0.8$. This distribution reflects the fact that two of the three axes should produce fairly flat (low $q$) shapes on the sky, while only projections nearly aligned with the prominent long axis should give projected shapes that are close to round ($q \sim 0.9$).  That is, round shapes are possible, but they should be rarer than flattened shapes if we are viewing an elongated shape from random directions.

\begin{figure*}
    \centering
    \includegraphics[width=1.0\linewidth]{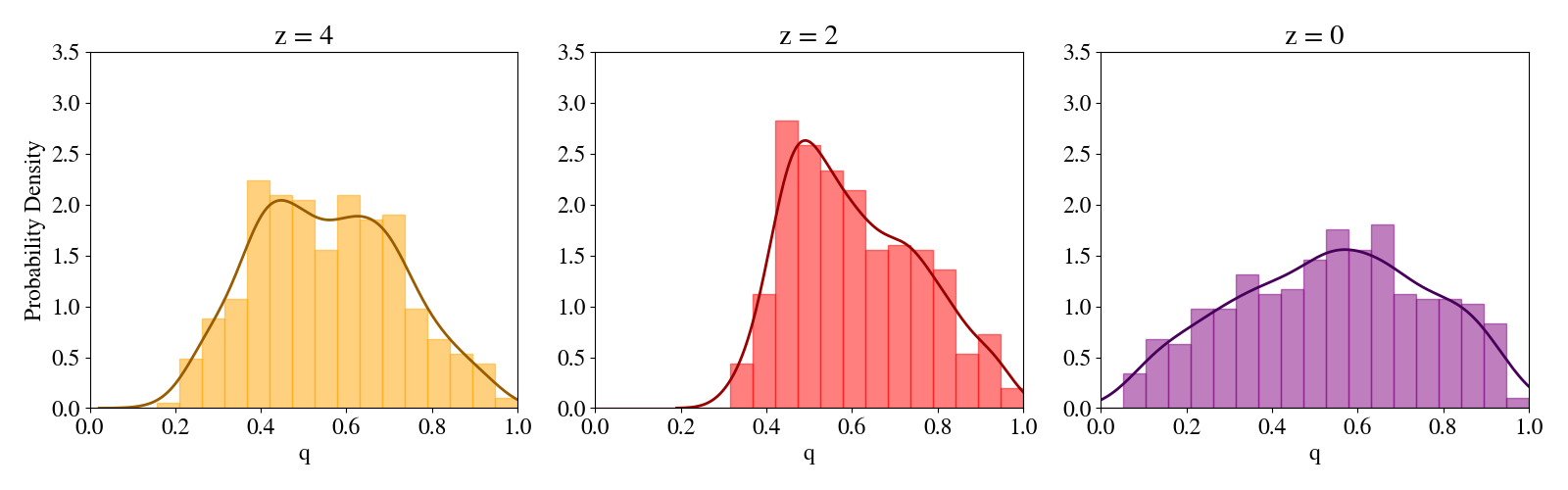}
    \caption{Predicted 2D shape distributions of FIRE-2 galaxies at $z = 0$, 2, and 4. Each histogram plots the density distribution of $q \equiv b/a$ at each redshift. The solid line is the KDE plot for each distribution. The $q$-value is obtained from fitting a S\'ersic brightness profile to mock images of randomly projected galaxies in our simulated sample. As the redshift decreases from 4 to 0, the FIRE-2 galaxy shapes transition from a spheroid-dominated distribution to a distribution that is relatively well-populated by disks ($q < 0.2$).}
    \label{fig:shape_distribution}
\end{figure*}

With these examples as context, Figure \ref{fig:shape_distribution} shows the predicted 2D shape distributions of all of the galaxies in our sample at $z = 0, 2,$ and $4$.  The ranges of stellar masses of the galaxies at these redshifts are approximately $10^{10-11}$, $10^{8.5 - 10}$, and $10^{7.5 - 9}$ M$_\odot$, respectively. Each of our 13 galaxies is projected along 30 random lines of sight at each redshift shown.  This gives 390 inferred $q$ values for each distribution.

When examining these three redshift distributions of projected shapes, it is useful to look at the ``Luminosity-Weighted" row in Figure ~\ref{fig:allgalaxies}, which shows the 3D shapes of the same galaxies at the same three redshifts.  From this figure, at $z = 0$, we know that the vast majority of our galaxies are disky in 3D. The 
$q$ distribution at $z=0$ shown in Figure \ref{fig:shape_distribution} is consistent with this fact.  Indeed, the full $z=0$ distribution is quite similar to the pure disky distribution shown in \ref{fig:pure_shape_distribution}: the tail  with $q < 0.2$ corresponds to edge-on views of disks and the tail with $q>0.8$ corresponds to face-on views. 

The 2D shape distribution at $z=2$ in Figure \ref{fig:shape_distribution} is more skewed than at $z=0$. This reflects the existence of a fairly large number of (luminosity-weighted) elongated galaxies at this redshift, as seen at $z=2$ in Figure~\ref{fig:allgalaxies}. The skewed peak we see in the full sample resembles the shape of the pure elongated sample from Figure \ref{fig:pure_shape_distribution}. A second, shallower peak at $q \sim 0.7-0.8$ is also discernible, suggesting a smaller spheroidal population.

Finally, the $z=4$ distribution in Figure \ref{fig:shape_distribution} shows a broader range of $q$ values than at $z=2$. This is consistent with the fact that there are fewer elongated systems and many more spheroidal shapes at $z=4$ in Figure~\ref{fig:allgalaxies}.

For comparison to observations, the best overlap between our stellar masses and redshifts and those of the \citet{Pandya2024} sample is at $z=2$ and $M_\star \simeq 10^{9-10}$M $_\odot$.  In this range, their Figure 1 shows distributions quite similar in shape to the middle panel of our Figure \ref{fig:shape_distribution}: skewed and peaked near $q \sim 0.45$ and with tails toward $q \sim 0.9$.  This is consistent with a sizable population of elongated galaxies in both the observed sample and our simulated sample.

\begin{figure}
    \centering
    \includegraphics[width=\columnwidth]{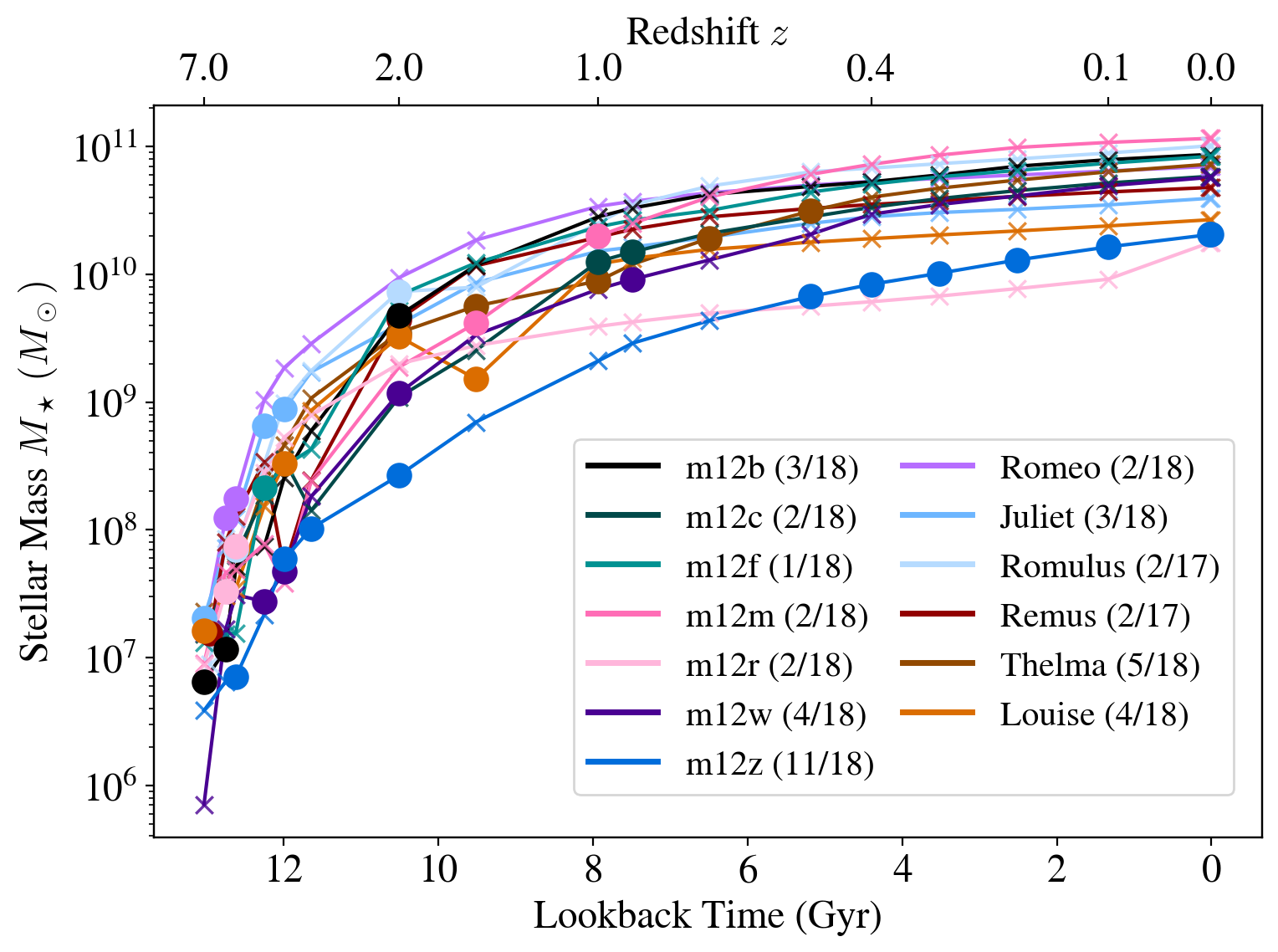}
    \caption{Stellar mass evolution of our 13 zoom-in galaxies as a function of lookback time. Snapshot timesteps where the shape of the young stars is elongated (as classified in Figure \ref{fig:baca_mapping}) are marked with circles.  The other timesteps are marked with x's. Each simulation name is color-coded as shown in the legend, along with the number of timesteps the galaxy is classified as elongated $e$ over the total number of timesteps $t$: ($e/t$).}
    \label{fig:mass_evolution}
\end{figure}

\begin{figure}
    \centering    
    \includegraphics[width=\columnwidth]{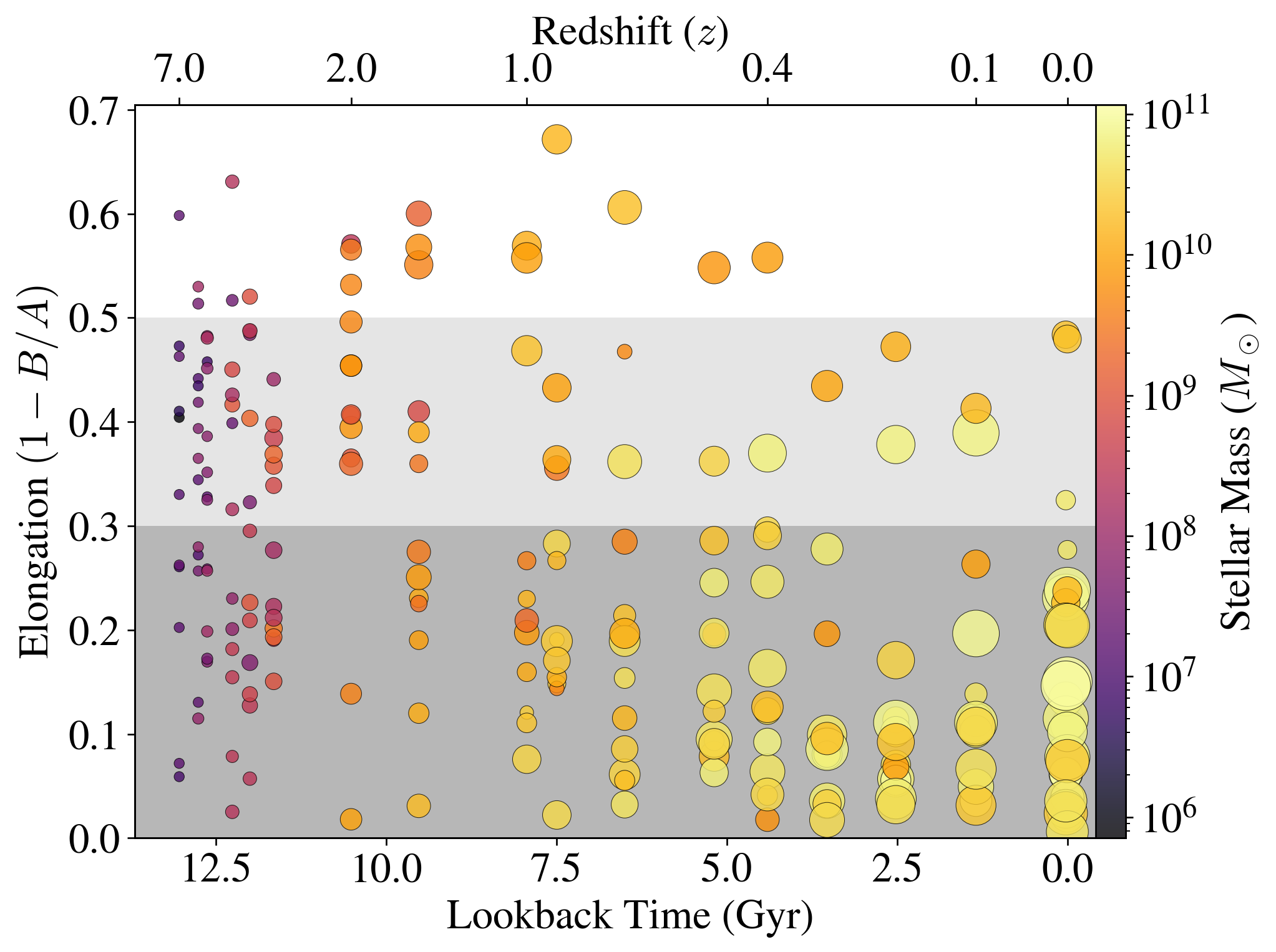}
    \caption{An elongation parameter, defined as \(1-B/A\) of the ellipsoid measured using young stars (\(<500\,\mathrm{Myr}\)), plotted against lookback time (with a redshift axis on top). The bands help to interpret the parameter as it relates to the ``elongated"  classification defined in Figure \ref{fig:baca_mapping}. The dark band (\(1-B/A = 0.0\text{--}0.3\)) corresponds to galaxies that are never classified as elongated.  The light band (\(0.3\text{--}0.5\)) includes values sometimes classified as elongated, depending on their $C/A$ ratios. The unshaded region (\(>\!0.5\)) contains values that are always classified as elongated. The marker size is proportional to the 3D galaxy radius,  defined as \(r_{90}\) of the stars remaining from the iterative reduced inertia tensor.   The smallest circle plotted corresponds to an \(r_{90}\) value of $0.68$ kpc and the largest circle corresponds to $34$ kpc.  The color of each circle represents the stellar mass of all stars. }
    \label{fig:elongation_time}
\end{figure}

\section{The Physical Properties of Galaxy Elongation}
\label{sec:elongation}
In this section we look at the physical properties associated with galaxy elongation to explore the origin of these shapes.  For much of the discussion that follows we utilize a crude measure of elongation ($\epsilon \equiv 1 - B/A$) using young stars ($< 500$ Myr). We also rely on the $n = 231$ snapshots that satisfy the particle-number criterion of Section~\ref{subsec:measure}, comprising 13 galaxies each with 17 or 18 snapshots.

\begin{figure*}
    \centering
    \includegraphics[width=1.0\linewidth]{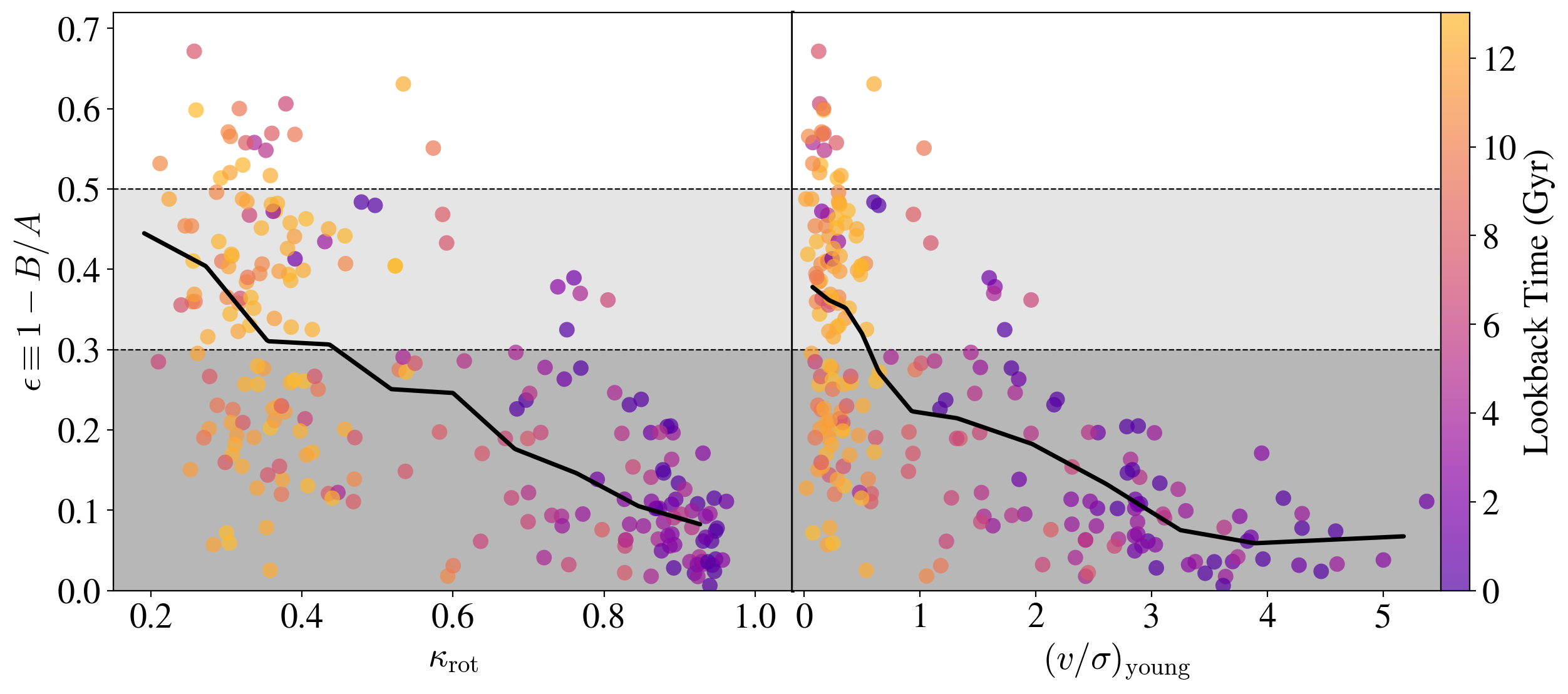}
    \caption{Rotational support of young stellar populations ($<500\,\mathrm{Myr}$) as a function of the elongation parameter $\epsilon \equiv 1 - B/A$, colored by lookback time. \textit{Left:} The fraction of kinetic energy in ordered rotation, $\kappa_{\rm rot}$, measured within $0.1\,r_{\rm vir}$. \textit{Right:} The rotational-to-dispersion ratio $(v/\sigma)_{\rm young}$, defined as the mean azimuthal velocity divided by the in-plane velocity dispersion. Vertical dashed lines and shading follow Figure~\ref{fig:elongation_time}. Elongated galaxies are systematically less rotationally supported, with Spearman rank correlations of $r = -0.64$ and $r = -0.67$ respectively.}
    \label{fig:kapparot}
\end{figure*}

\subsection{Transient Elongation and Evolution with Mass and Size}

Figure \ref{fig:mass_evolution} plots the stellar mass evolution of each of our simulated galaxies, each of which has either 17 or 18 snapshots.  Each snapshot is marked by either a filled circle or an x. The filled circles indicate that the galaxy has a young-star population that is elongated according to the classification in Figure \ref{fig:baca_mapping}.  All other timesteps are marked as x's. 

We see that every galaxy experiences periods when its young stars have an elongated shape.  The fraction of timesteps where we measure the young stars to be elongated ranges from $\sim 6\%$ (m12f) to $\sim 60\%$ (m12z). In several cases, one timestep is elongated (solid circle) while the 
next is not (x's), showing that these phases are often short-lived, 
lasting $\lesssim 2$ Gyr.  A notable exception is m12z, which experiences a series of fairly significant mergers from a lookback time $\sim 6$ Gyr down to $z=0$.~\footnote{Note that even in this case, the orientations of the measured elongated ellipsoids for m12z at late times are not the same, suggesting that the shapes are not stable.} 

Taken as an ensemble, our galaxies experience elongated phases over a range of stellar masses, from $\sim 10^{6.5-7.5}$ M$_\odot$ (at $z \sim 5-7$) to $\sim 10^{9.5-10.5}$ M$_\odot$ (at $z\sim 0-1.5$), though the fraction is smaller at lower redshift.

Figure \ref{fig:elongation_time} gives another view of our ensemble results, now using our $\epsilon \equiv 1 - B/A$ parameter to track elongation as a function of lookback time. Circles show $\epsilon$ measured from young stars at each snapshot. The color of the circle maps to the galaxy stellar mass, as indicated in the color bar. Circle size is proportional to the $r_{90}$ radius of the stars in the galaxy.

As defined in Figure \ref{fig:baca_mapping}, whether a galaxy is classified as elongated depends both on $B/A$ and $C/A$, but it is always the case that when $1-B/A$ is greater than $0.5$ galaxies are elongated. This corresponds to the non-shaded region in Figure \ref{fig:elongation_time}.   For intermediate values, $1-B/A = 0.3 - 0.5$, galaxies are sometimes elongated and sometimes spheroidal or disky, depending on $C/A$. This region has medium shading.  For  $1-B/A < 0.3$, the galaxy is never elongated.  This region is shaded dark. 

Figure \ref{fig:elongation_time} shows that at every redshift, at least some of our sample galaxies are elongated.  However,  the most extreme examples ($1-B/A > 0.5$) for massive ($\gtrsim 10^{9.5}$ M$_\odot$) systems occur between $z=2$ and $z=0.4$. Finally, we note that no elongated shapes are seen after a lookback time of $\sim 6$ Gyr, except for m12z, which undergoes a series of mergers.  The rest of the sample transitions to a rotating thick disk or thin disk phase after this time  \citep[see][]{2023MNRAS.523.6220Y}.

\subsection{Stellar Kinematics}
\label{sec:kinematics}

Next, we examine how young-star elongation correlates with the kinematic properties of the young stellar population. First, we calculate $\kappa_{\mathrm{rot}}$, the rotational support parameter used by \citet{2012MNRAS.423.1544S} and \citet{2025MNRAS.544.4651B}. This is defined as the fraction of kinetic energy in ordered rotation relative to the total stellar kinetic energy:
\begin{equation}
\kappa_{\mathrm{rot}} = \frac{K_{\mathrm{rot}}}{K}
= \frac{1}{K} \sum \frac{1}{2} m \left( \frac{j_z}{R} \right)^2,
\end{equation}
where $K$ is the total kinetic energy of stellar particles within $r_{\mathrm{gal}}$, $j_z$ is the angular momentum component along the net angular momentum axis (after alignment with the $z$-axis), $m$ is the particle mass, and $R$ is the cylindrical radius. The sum is taken over the young ($< 500$ Myr) stellar particles within $0.1\,r_{\mathrm{vir}}$. High values of $\kappa_{\mathrm{rot}} \gtrsim 0.5$ indicate rotation-supported (disk-like) systems, while low values $\kappa_{\mathrm{rot}} \lesssim 0.35$ correspond to dispersion-dominated systems, with intermediate values reflecting mixed kinematics. We also measure $(v/\sigma)_{\rm young}$ for the same stellar population, defined as the ratio of the mass-weighted mean azimuthal velocity $\langle v_\phi \rangle$ to the one-dimensional velocity dispersion $\sigma_{\rm 1d} \equiv \sqrt{(\sigma_R^2 + \sigma_\phi^2)/2}$, where $\sigma_R$ and $\sigma_\phi$ are the mass-weighted radial and azimuthal dispersions computed in the angular momentum frame within $0.1\,r_{\rm vir}$.

Figure~\ref{fig:kapparot} shows $\epsilon$ as a function of both quantities. Elongated systems are systematically less rotationally supported (see caption for Spearman correlations). The takeaway is clear: elongated configurations are dispersion-dominated rather than rotation-supported.

To characterize the anisotropy of the velocity dispersion, we compute the mass-weighted velocity dispersion tensor in the shape eigenbasis established by the reduced inertia tensor: \begin{equation} \Sigma_{ij} = \frac{\sum_n m_n \,\delta v_i^{(n)}\, \delta v_j^{(n)}}{\sum_n m_n}, \label{eq:vel_tensor} \end{equation} where $\delta \mathbf{v}^{(n)} = \mathbf{v}^{(n)} - \bar{\mathbf{v}}$ is the peculiar velocity of the $n$-th star particle relative to the mass-weighted mean $\bar{\mathbf{v}}$, and the indices $i, j$ run over the three spatial dimensions in the shape frame. Diagonalizing $\Sigma_{ij}$ yields eigenvalues whose square roots we label $\sigma_1 \geq \sigma_2 \geq \sigma_3$, corresponding to the dispersions along the major, intermediate, and minor axes of the spatial distribution ($\hat{A}$, $\hat{B}$, $\hat{C}$) respectively.

\begin{figure*}
    \centering
    \includegraphics[width=1.0\linewidth]{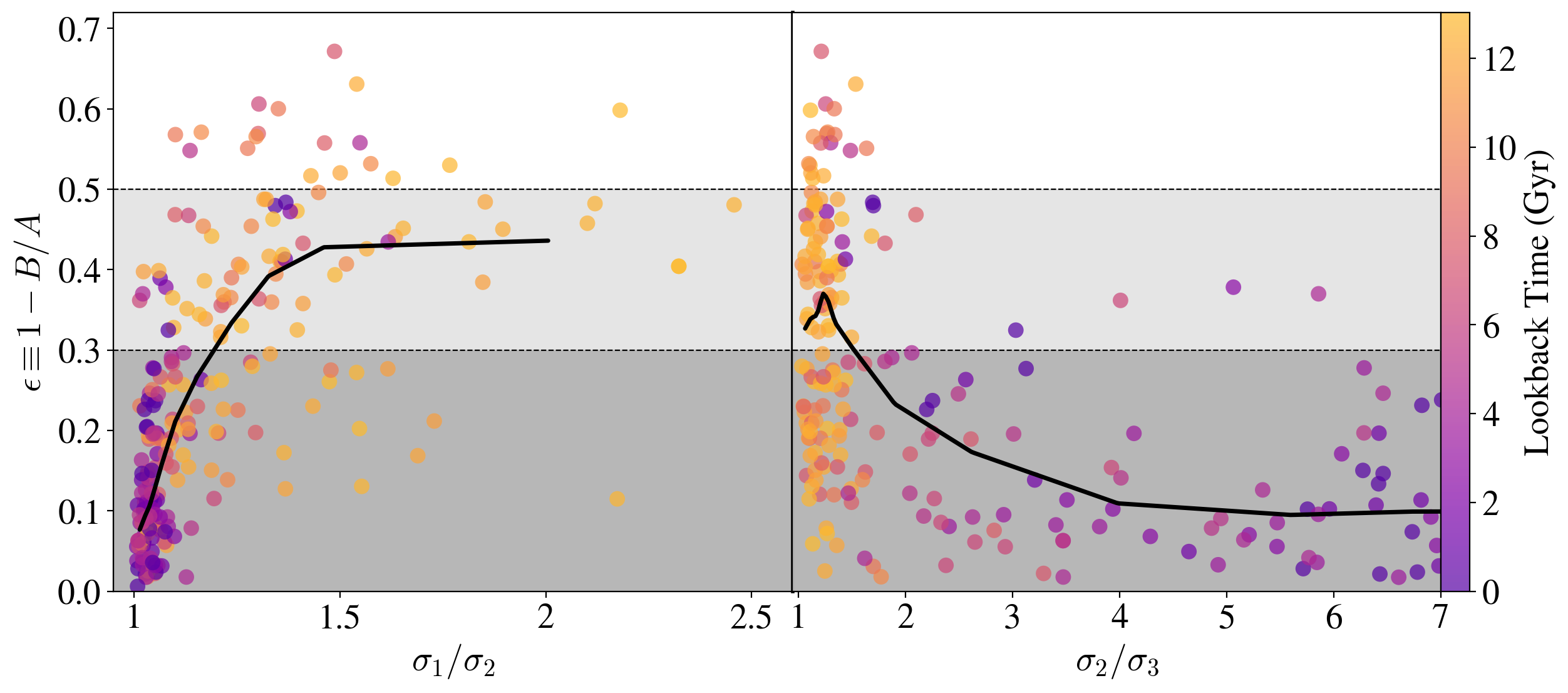}
    \caption{Velocity ellipsoid shape ratios for young stellar populations ($<500\,\mathrm{Myr}$) as a function of the elongation parameter $\epsilon \equiv 1 - B/A$, colored by lookback time. $\sigma_1 \geq \sigma_2 \geq \sigma_3$ are the eigenvalues of the velocity dispersion tensor computed in the shape frame, ordered from largest to smallest. \textit{Left:} $\sigma_1/\sigma_2$ increases strongly with elongation (Spearman $r = +0.72$), indicating that the dominant dispersion direction separates from the intermediate. \textit{Right:} $\sigma_2/\sigma_3$ decreases with elongation ($r = -0.58$), showing that the intermediate and minor dispersion components converge. Together, non-elongated systems have oblate velocity ellipsoids ($\sigma_1 \sim \sigma_2 \gg \sigma_3$; mean $\sigma_1/\sigma_2 = 1.11$, $\sigma_2/\sigma_3 = 3.60$), while elongated systems have prolate velocity ellipsoids ($\sigma_1 \gg \sigma_2 \sim \sigma_3$; mean $\sigma_1/\sigma_2 = 1.45$, $\sigma_2/\sigma_3 = 1.27$). The solid black line shows the running median in equal-count bins. Vertical dashed lines and shading follow Figure~\ref{fig:elongation_time}.}
    \label{fig:vel_ellipsoid}
\end{figure*}

Figure~\ref{fig:vel_ellipsoid} shows the velocity ellipsoid ratios $\sigma_1/\sigma_2$ and $\sigma_2/\sigma_3$ as a function of $\epsilon$, where $\sigma_1 \geq \sigma_2 \geq \sigma_3$ are the eigenvalues of the velocity dispersion tensor computed in the shape frame. Non-elongated systems have oblate velocity ellipsoids ($\sigma_1 \sim \sigma_2 \gg \sigma_3$), consistent with rotation-supported disks. In contrast, elongated systems have prolate velocity ellipsoids ($\sigma_1 \gg \sigma_2 \sim \sigma_3$), where the dominant dispersion is concentrated along a single preferred axis.

\begin{figure*}
    \includegraphics[width=1.0\linewidth]{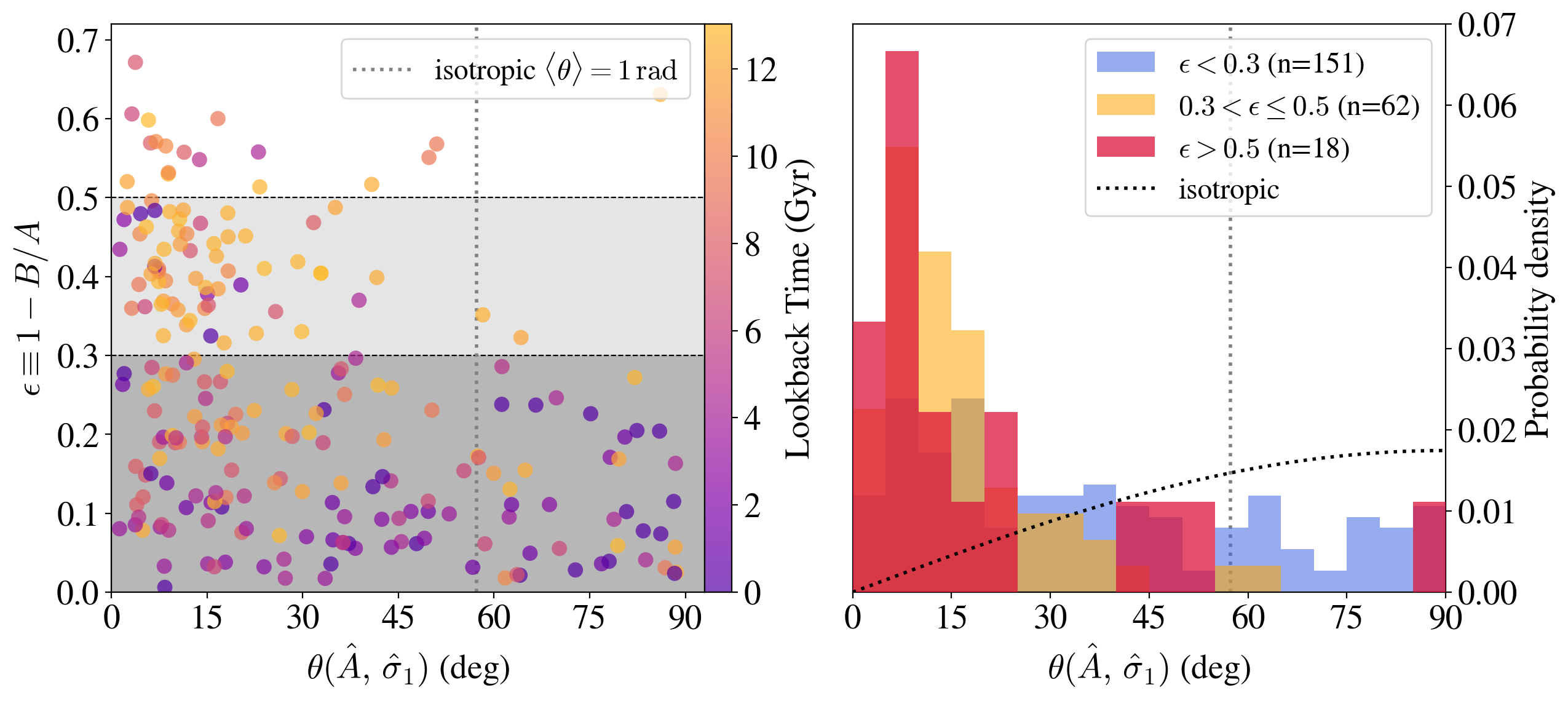}
    \caption{Alignment between the spatial and kinematic major axes of young stellar populations ($<500\,\mathrm{Myr}$). \textit{Left:} The angle $\theta$ between the shape eigenvector $\hat{A}$ (direction of maximum spatial extent) and the velocity dispersion eigenvector $\hat{\sigma}_1$ (direction of maximum dispersion), plotted against $\epsilon \equiv 1 - B/A$ and colored by lookback time. The dotted horizontal line marks the isotropic expectation of $57.3\degr$. Vertical dashed lines and shading follow Figure~\ref{fig:elongation_time}. \textit{Right:} Normalized distributions of $\theta$ in three elongation bins. elongated systems ($\epsilon > 0.5$) show strong alignment, with 74 percent of snapshots within $20\degr$ and a median of $8.6\degr$, compared to a median of $30.3\degr$ for non-elongated systems ($\epsilon < 0.3$). A KS test rejects the null hypothesis that the two distributions are the same ($p = 2.4 \times 10^{-5}$). The dotted curve shows the isotropic expectation $p(\theta) \propto \sin\theta$.}
    \label{fig:axis_alignment}
\end{figure*}

This naturally raises the question of whether that kinematic axis aligns with the spatial major axis. Figure~\ref{fig:axis_alignment} shows the angle $\theta(\hat{A}, \hat{\sigma}_1)$ between the shape eigenvector $\hat{A}$ and the velocity dispersion eigenvector $\hat{\sigma}_1$ as a function of $\epsilon$. During elongated phases the two axes are strongly aligned, with a median alignment angle of just $8.6^\circ$ compared to $30.3^\circ$ for non-elongated systems; a KS test rejects the null hypothesis that the two distributions are drawn from the same population ($p = 2.4 \times 10^{-5}$). Taken together, Figures~\ref{fig:kapparot}--\ref{fig:axis_alignment} paint a consistent picture: elongated galaxies have an anisotropic velocity dispersion directed along their spatial major axis, a signature qualitatively similar to what \citet{Tomassetti2016} found in the Vela simulations. As we discuss in Section~\ref{sec:dm}, however, the dark matter properties of elongated systems in our FIRE-2 sample differ substantially from the Vela predictions, suggesting that this kinematic state can arise through different physical pathways.

\subsection{Gas Kinematics and Filamentary Accretion}
\label{sec:filamentary}

\citet{Tomassetti2016} suggested that elongated stellar configurations in their Vela simulations were driven by anisotropic accretion along large-scale filaments, with DM halo torques supporting the elongated stellar shape and aligning it with the filament direction \citep{Ceverino15,2023MNRAS.522.4515L}. 

In this subsection we explore whether or not  halo-scale gas inflow aligns with the stellar major axis during elongated phases. For each snapshot, we define an inflow axis $\hat{\mathbf{A}}_{\rm inflow}$ from the flux-weighted dipole of gas particles inflowing through a spherical shell $r \in [r_1, r_2]$. 
\begin{equation}
\mathbf{F}_{\rm inflow} = \sum_{v_{r,i}<0} m_i\,(-v_{r,i})\,\hat{\mathbf{r}}_i, \qquad
\hat{\mathbf{A}}_{\rm inflow} = \frac{\mathbf{F}_{\rm inflow}}{|\mathbf{F}_{\rm inflow}|}.
\end{equation}
Note that $\mathbf{F}_{\rm inflow}$ is proportional to the first angular moment of the inward mass flux across the shell, and when inflow is dominated by a single coherent stream, $\hat{\mathbf{A}}_{\rm inflow}$ points along the direction of maximum mass inflow; when inflow is more isotropic or multi-stream, the dipole amplitude is reduced. We quantify the coherence of the inflow field using

\begin{equation}
\mathcal{C}_{\rm inflow} = \frac{|\mathbf{F}_{\rm inflow}|}{\sum_{v_{r,i}<0} m_i\,(-v_{r,i})},
\end{equation}
and restrict alignment measurements to snapshots with sufficient inflowing mass and $\mathcal{C}_{\rm inflow}$ above a minimal threshold. 

To capture cases where opposite streams cancel in the dipole moment, 
we also define a quadrupole inflow axis $\hat{\mathbf{A}}_{\rm inflow}^{(2)}$ 
as the principal eigenvector of the flux-weighted tensor
\begin{equation}
\mathbf{M}_{\rm inflow} \equiv 
\frac{\sum_{v_{r,i}<0} m_i\,(-v_{r,i})\,\hat{\mathbf{r}}_i\hat{\mathbf{r}}_i^{\rm T}}
{\sum_{v_{r,i}<0} m_i\,(-v_{r,i})},
\end{equation}
with anisotropy quantified by $\mathcal{C}_{\rm quad} \equiv 
(\lambda_1-\lambda_3)/(\lambda_1+\lambda_2+\lambda_3)$, which ranges from 0 for isotropic inflow to 1 for maximally anisotropic inflow. We then measure the alignment of both inflow axes with the stellar major axis:
\begin{equation}
\theta_{\rm fil} = \arccos\left(|\hat{A}\cdot\hat{A}_{\rm inflow}|\right), 
\quad 
\theta_{\rm fil}^{(2)}=\arccos\left(|\hat{A}\cdot\hat{A}_{\rm inflow}^{(2)}|\right).
\end{equation}







Despite testing shell sizes of $r \in [0.1, 1.0]\,r_{\rm vir}, \,r \in [1.0, 2.0]\,r_{\rm vir}$, coherence parameter cuts at $\mathcal{C}_{\rm inflow}= 0.5, 0.7, 0.9$, and even allowing a time lag of 1-2 snapshots, we find no statistically significant correlation between either alignment metric and the stellar shape at any elongation value or redshift ($\text{Spearman }| r| < 0.2, \, p>0.05$, in all cases). 

There are at least two reasons this null result may still allow preferential accretion to play a role in elongation. First, our snapshot spacing of $\sim 750$~Myr is likely too coarse to capture the relevant signal. \citet{Sultan2026} find that in bursty galaxies, the regime applicable to our high-$z$ progenitors prior to inner CGM virialization, accreted cold gas typically forms stars after only $\lesssim 5$ galaxy free-fall times (their Figure~14). For Milky Way progenitor masses at relevant redshifts, this corresponds to only $\sim 250$~Myr, which is a third of our temporal resolution. Additionally, \citet{Sultan2026} show that cold gas in pre-virialized halos is morphologically complex and disordered rather than arriving in narrow coherent streams.  A dedicated analysis with finer time resolution and particle tracking would be required to test the filamentary hypothesis properly.

\subsection{Dark Matter Shapes}
\label{sec:dm}
\citet{Ceverino15} and \citet{Tomassetti2016} found that elongated phases in their Vela simulations coincided with dark-matter-dominated systems whose halos were themselves elongated and aligned with the stellar distribution, suggesting that anisotropic assembly along large-scale filaments drives stellar elongation. We test whether similar trends hold in our FIRE-2 sample.

\begin{figure}
    \centering
    \includegraphics[width=\columnwidth]{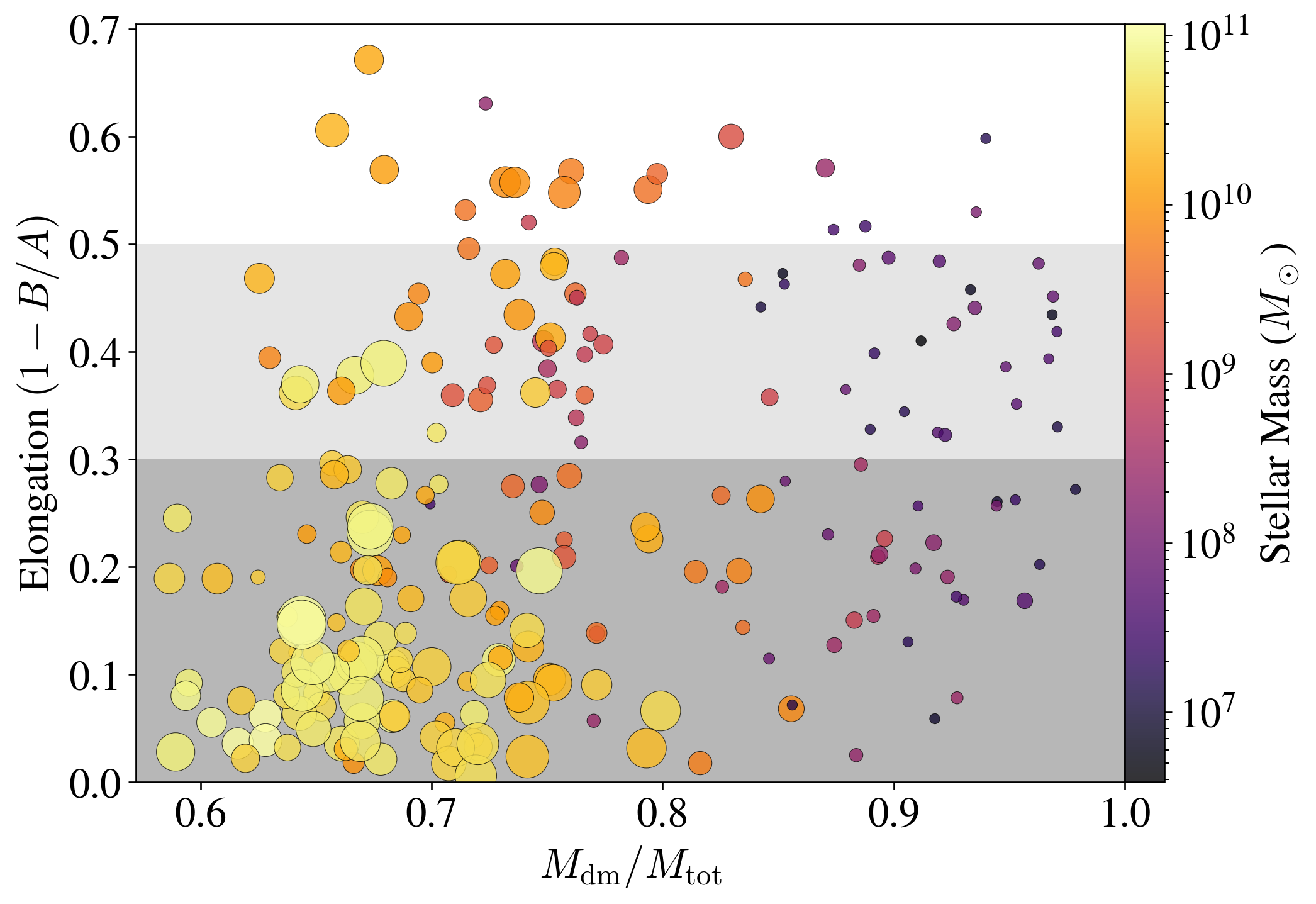}
    \caption{Elongation $\epsilon \equiv 1 - B/A$ of young stars as a function of the dark matter mass fraction $M_{\rm dm}/M_{\rm tot}$ within $0.1\,r_{\rm vir}$. Colors, sizes, and shading follow Figure~\ref{fig:elongation_time}. There is no clear trend with dark matter fraction.}
    \label{fig:dm_frac}
\end{figure}

Figure~\ref{fig:dm_frac} shows elongation as a function of dark matter mass fraction within $0.1\,r_{\rm vir}$. We find no clear correlation: elongated and non-elongated systems span the same range of dark matter fractions at all redshifts. 

\begin{figure*}
    \includegraphics[width=\textwidth]{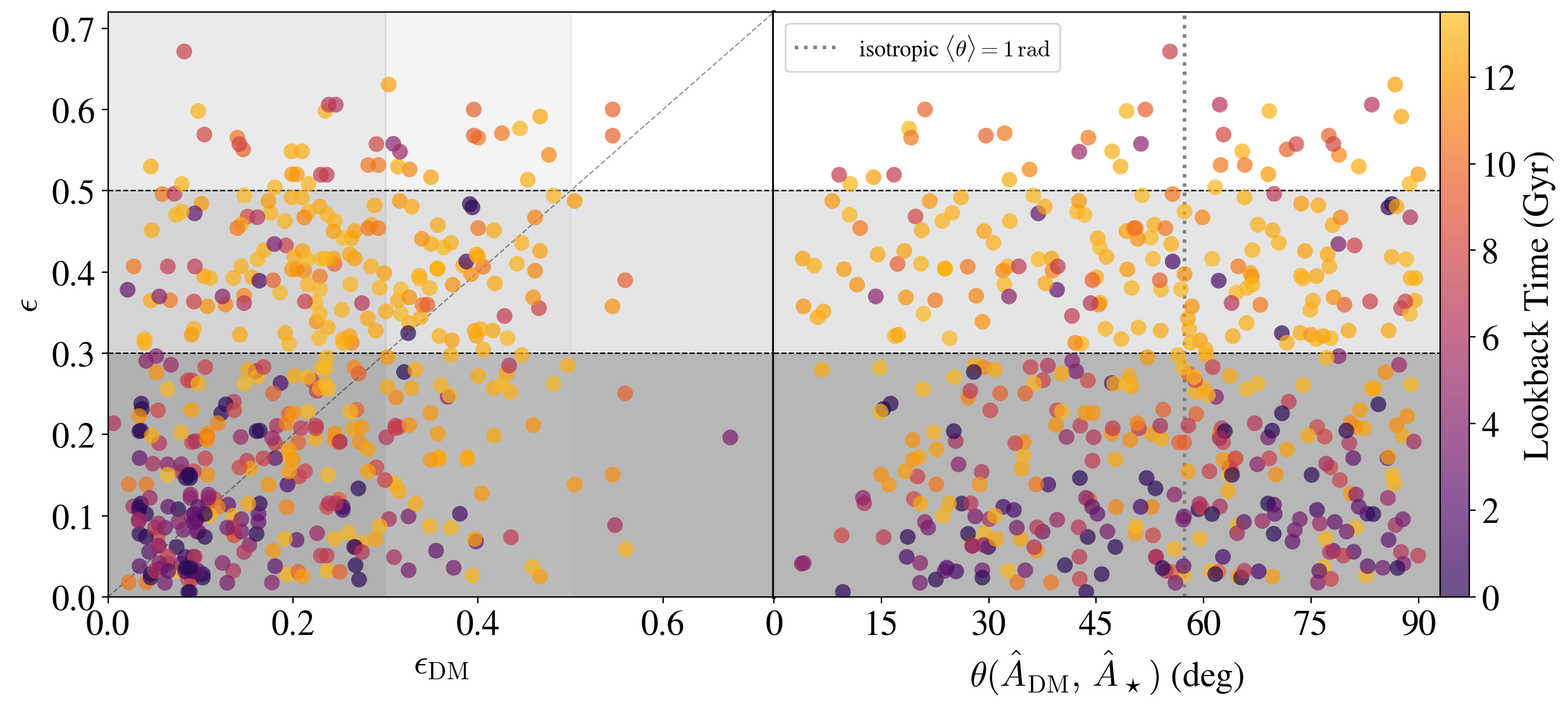}
    \caption{Dark matter halo properties as a function of young-star elongation $\epsilon_{\rm young} \equiv 1 - B/A$, colored by redshift. \textit{Left:} Dark matter halo elongation $\epsilon_{\rm DM} \equiv 1 - B/A$ measured within $r_{\rm vir}$; the dashed line shows the 1:1 relation. Dark matter halos are systematically rounder than the stellar component at all epochs (Spearman $r = +0.31$). \textit{Right:} Alignment angle between the dark matter major axis $\hat{A}_{\rm DM}$ and the young-star major axis $\hat{A}_{\rm young}$. The dotted line marks the isotropic expectation of $57.3^{\circ}$. The orientation of the dark matter halo is consistent with random at all elongation values (Spearman $r = +0.004$). Shading follows Figure~\ref{fig:elongation_time}.}
    \label{fig:dm_shape}
\end{figure*}

Figure~\ref{fig:dm_shape} reinforces this picture. The dark matter halo is marginally more elongated during elongated stellar phases, but the effect is weak (Spearman $r = +0.31$). Here the dark matter shape is measured within $r_{\rm vir}$; the correlation is fully absent when measured within $0.1\,r_{\rm vir}$ (because the dark matter is almost spherical in the inner region). Furthermore, the orientation of the dark matter major axis bears no relationship to the stellar major axis at any elongation value (Spearman $r = +0.004$), contrary to past work. Together, these results suggest that the elongation mechanism in FIRE-2 is not driven by dark matter co-elongation or alignment, in contrast to the Vela predictions. One potential explanation is that the stronger stellar feedback implementation in FIRE-2 more efficiently redistributes baryonic mass and disrupts the correspondence between the inner stellar distribution and the large-scale dark matter structure. In this picture, feedback-driven outflows and turbulent gas motions may generate elongated stellar configurations through internal dynamical processes rather than through alignment with an elongated host halo.

\section{Discussion and Conclusions}
\label{sec:conclusions}
We have studied the 3D structure of Milky Way–like galaxies simulated with FIRE-2 physics using a triaxial ellipsoid approximation to characterize their shapes. Of the thirteen galaxies in our sample, twelve form disks at $z=0$.  The one exception (m12z) has a series of mergers at late times. We refer to the non-merging galaxies as ``Milky Way analogs."

All of our galaxies go through phases in the early Universe when they were either spheroidal or elongated (like a pickle).  During these early periods, their relative axis ratios vary significantly over Gyr timescales and often transition from spheroidal to elongated and back again, though there is significant variance (see Figures \ref{fig:Thelmapoints}, \ref{fig:mass_evolution}, and \ref{allm12Lumweight}). We find that these general trends hold whether we measure shapes weighted by stellar mass or stellar luminosity, though the elongation is more significant when weighted by luminosity.  

We also studied how the shapes of stellar populations change from the time of their formation to their final configurations at $z=0$. As illustrated in Figures \ref{fig:Thelma_pastpresent}, \ref{fig:thelmaevolution}, and \ref{allm12Archaeology}, stellar populations at $z=0$ have shapes that change systematically with age: the youngest stars inhabit thin disks, intermediate-age stars have thick-disk configurations, and the oldest stars reside in flattened spheroids.  In all of our Milky Way analogs, the $z=0$ populations are symmetric about their minor axes, with intermediate-to-major-axis ratios $B/A \sim 0.9$. These $z=0$ shapes are drastically different from the shapes those same stellar populations had at the time of their formation (Figures \ref{fig:Thelmapoints} and \ref{allm12YoungStars}).  In particular, early-forming populations tended to have either triaxial-spheroidal or elongated shapes at birth. The implication is that these populations change shape significantly over cosmic time. Early elongated configurations, in particular, are not stable over a Hubble time.

Although our analysis traces the evolution of a specific set of $z=0$ Milky Way-like galaxies back in time, and therefore does not represent a fair cosmic sample, we did make a cursory comparison to the observed shape distributions reported in \citet{Pandya2024}. Figure~\ref{fig:shape_fractions}, for example, shows that $\sim 35\%$ of our galaxies at $z \sim 2$ are elongated in shape, which is similar to the $\sim 40\%$ fraction reported by \citet{Pandya2024} for their similar-mass galaxies at the same redshift. We also showed using mock observations that our 2D axis ratio distributions on the sky at $z=2$ (see Figure \ref{fig:shape_distribution}) are quite similar to the distributions found at the same mass and redshift in \citet{Pandya2024} Figure 1. 

Having established that elongated galaxies exist in both the real universe and our simulations, we now summarize the physical picture that emerges. We find that elongated (spatially prolate) systems tend to lack significant coherent rotation (Figure \ref{fig:kapparot}) and also have prolate velocity ellipsoids ($\sigma_1 \gg \sigma_2 \sim \sigma_3$) with the dominant dispersion oriented along the long spatial axis (Figures \ref{fig:vel_ellipsoid} and \ref{fig:axis_alignment}). 

We next examined correlations between the dark matter and stellar shapes in our galaxies.  This question is motivated by the work of \citet{Ceverino15} and \citet{Tomassetti2016}, who also found that their simulated galaxies underwent transient elongation phases.  In their Vela simulations, they found that elongation occurred when  their galaxies were dark-matter dominated and when the dark-matter halo was also elongated and aligned with the stars. In contrast, we find that the elongation in our galaxies is {\em not} correlated with either their dark matter fraction (Figure \ref{fig:dm_frac}) or the shape of their dark matter halo (Figure \ref{fig:dm_shape}). We also find that the long axes of our elongated galaxies are {\em not} correlated with any large-scale flow axis that we could find within the timestep limitations of our simulation outputs ($\sim 750$Myr).  

Despite several null results reported here, an important goal going forward is to understand the physical drivers of elongation in galaxies, and to ask whether those drivers are always the same from galaxy to galaxy.  Both our work and previous work by \citet{Ceverino15} and \citet{Tomassetti2016} found that the elongation phase tends to be transient in time for each galaxy. However, in our simulations we see that the fraction of galaxies that are elongated at any given time is sizable, suggesting that the duty cycles for triggering elongation and for relaxation back to more symmetric shapes are comparable. Mergers and trigger-aligned star formation may play a role.  In this vein, the only galaxy in our sample that persists as elongated at late times is m12z, which experiences a series of late-time mergers. We also find a few elongated systems with non-trivial rotational support ($\kappa_{\rm rot} \sim 0.6$); these correspond to populations in Figure~\ref{fig:Thelma_pastpresent} that were born with elongated configurations and that ultimately settle into thick disks rather than spheroids by $z=0$.

Connecting $z=0$ stellar kinematics to birth structures in the Milky Way is a promising avenue for future work. It is well known that 6D phase-space plus chemical information in the Milky Way can provide a powerful way to trace back the origins of its subcomponents \citep[e.g.][]{2024NewAR..9901706D}. High-dimensional phase-space structures may be a window into the past shapes of Galactic stellar populations, which could then enable a direct connection to the elongated shapes we observe in distant galaxies billions of years ago.

\section*{Acknowledgments}
LYX, JSB, and JDW are supported by NSF grant AST-1910965 and NASA grant 80NSSC22K0827. CK is supported by a National Science Foundation (NSF) Graduate Research Fellowship Program under grant DGE-1839285. MBK acknowledges support from NSF grants AST-1910346, AST-2108962, and AST-2408247; NASA grant 80NSSC22K0827; HST-GO-16686, HST-AR-17028, HST-AR-17043, JWST-GO-03788, and JWST-AR-06278 from the Space Telescope Science Institute, which is operated by AURA, Inc., under NASA contract NAS5-26555; and from the Samuel T. and Fern Yanagisawa Regents Professorship in Astronomy at UT Austin. LYX thanks CK and JSB for their stellar mentorship. LYX also thanks Emmeline Kim, Bryan Nnadi, James Buda, Marco Cheng, and Lily Chen for their comments and suggestions.

\section*{Data Availability}
The data underlying the figures in this paper are available via the online repository Zenodo at \href{https://doi.org/10.5281/zenodo.19850804}{doi:10.5281/zenodo.19850804}.

\bibliographystyle{mnras}
\bibliography{bib} 
\clearpage

\appendix

\section{Measurements of the Other 12 Galaxies}
\label{sec:appendix2}
\begin{figure*}
    \centering
    \includegraphics[width=\textwidth]{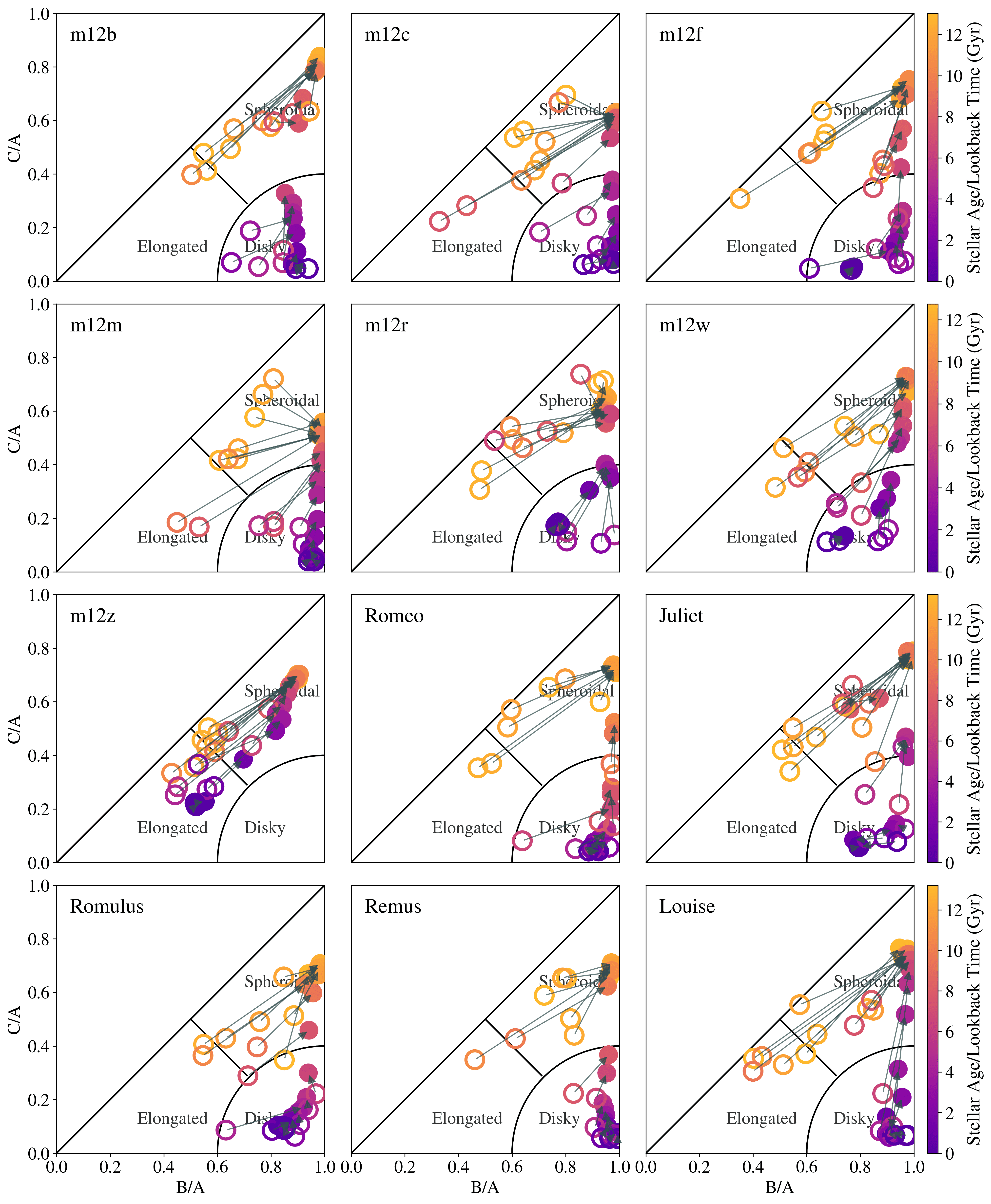}
    \caption{This figure uses the parameter space introduced in Figure \ref{fig:baca_mapping} and mimics the presentation provided for Thelma in Figure \ref{fig:Thelma_pastpresent} for the remaining 12 galaxies in our sample. The solid points depict the axis ratios of $z=0$ stellar populations binned by age (color bar).  The open circles show the shapes of those stellar populations at their formation times. We see evolution towards symmetry about the minor axis ($B/A \sim 1$) in all cases except for m12z, which is experiencing a major merger at $z=0$.} 
    \label{allgalaxiespastpresent}
\end{figure*}

\begin{figure*}
    \centering
    \includegraphics[width=\textwidth]{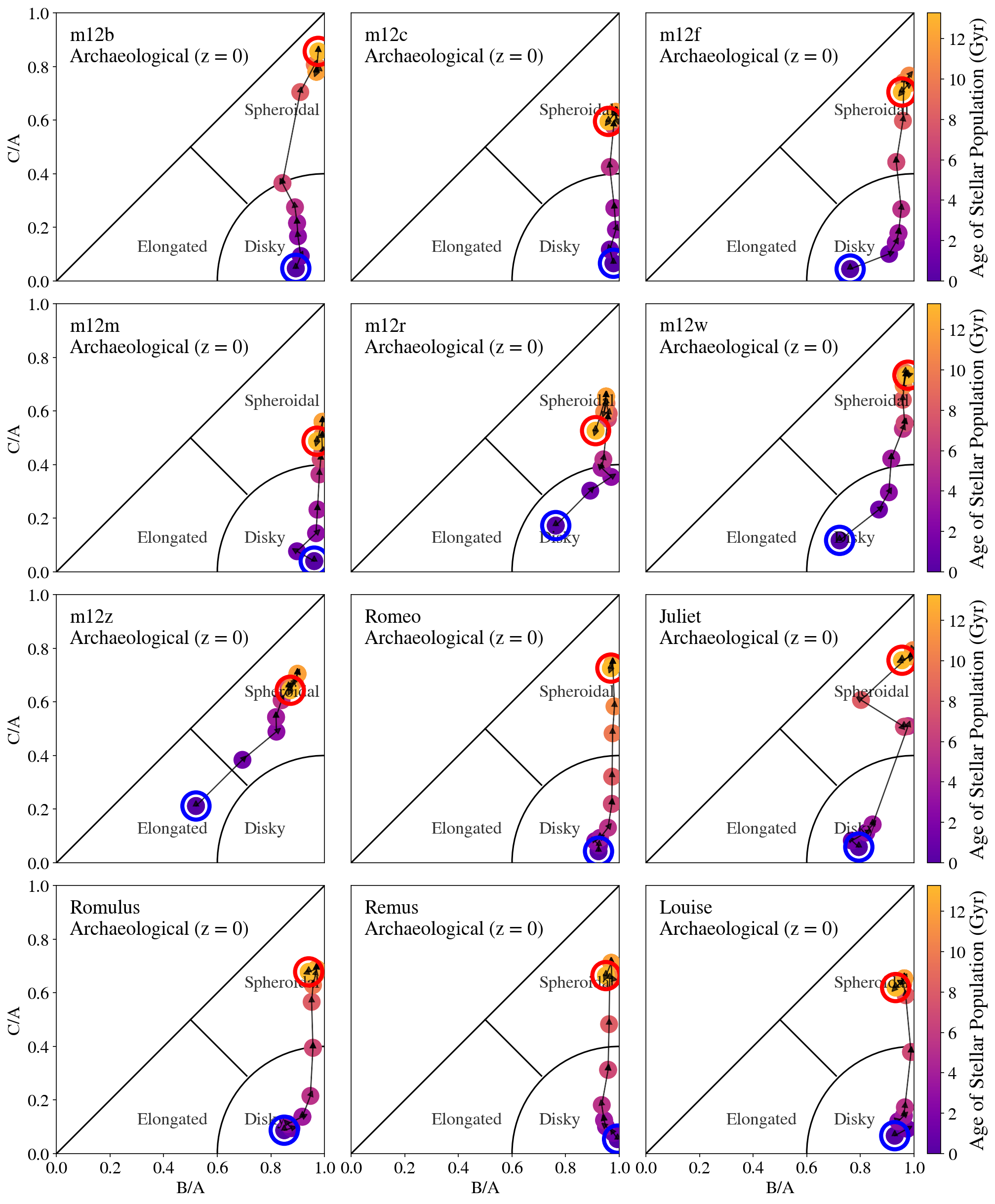}
    \caption{Archaeological decomposition of simulated galaxies, mimicking the top left panel of Figure \ref{fig:Thelmapoints}. The arrows connect points in time (color bar), starting with the earliest timestep. With the exception of m12z, which is undergoing a merger, they all exhibit similar behavior to Thelma in Figure \ref{fig:Thelmapoints}: stars inhabit progressively thinner disks (progressively smaller $C/A$) as the populations become younger.  All populations have symmetric shapes about their minor axes (high $B/A$ ratios).}
    
    \label{allm12Archaeology}
\end{figure*}

\begin{figure*}
    \centering
    \includegraphics[width=\textwidth]{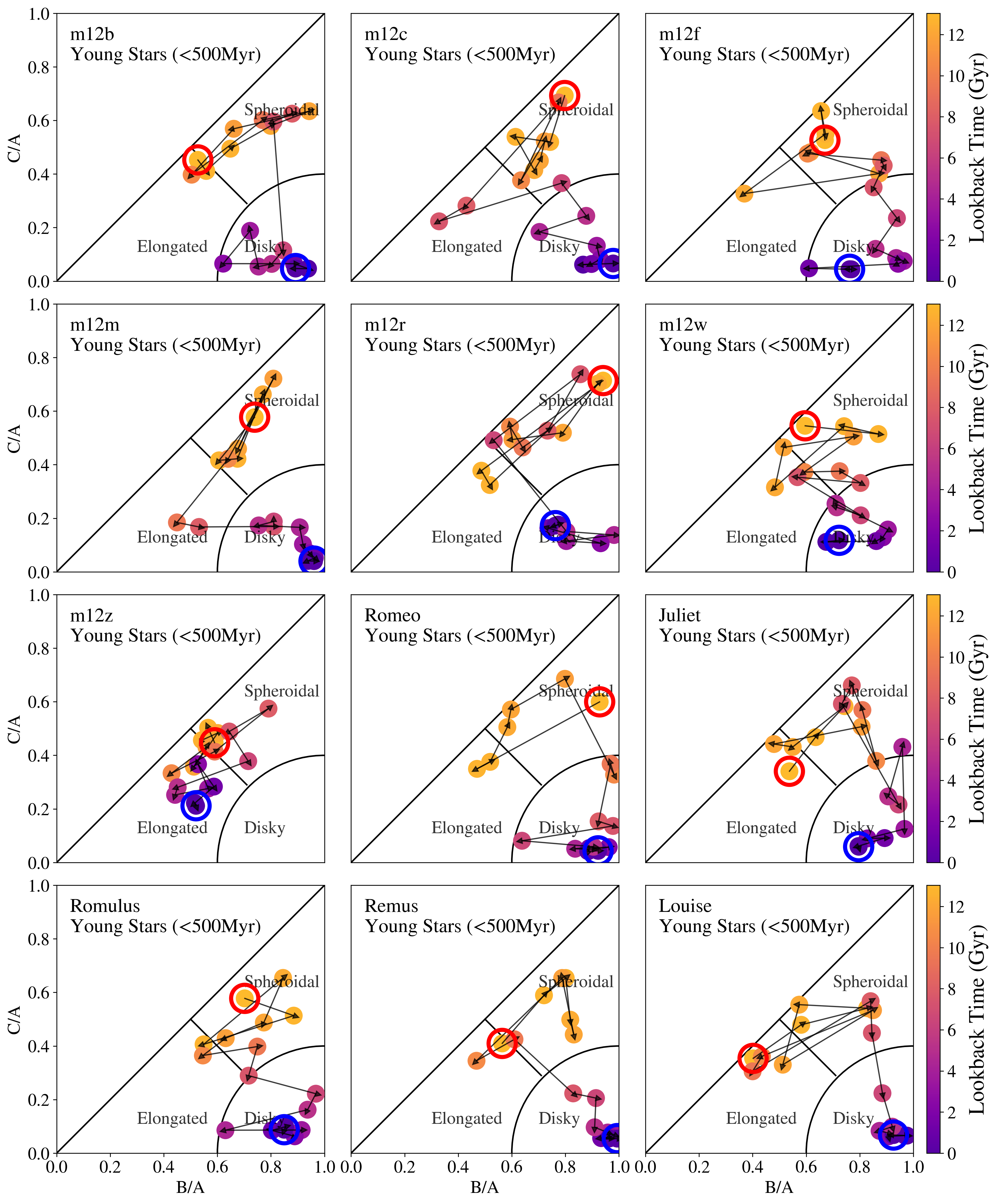}
    \caption{The young-stars shape evolution of simulated galaxies, corresponding to the top right panel of Figure \ref{fig:Thelmapoints}. Every galaxy goes through an elongated phase and a spheroidal phase at some point in the early Universe. In every case except m12z, which undergoes a late-time merger, young stars form disks as we approach the present day.}
    \label{allm12YoungStars}
\end{figure*}

\begin{figure*}
    \centering
    \includegraphics[width=\textwidth]{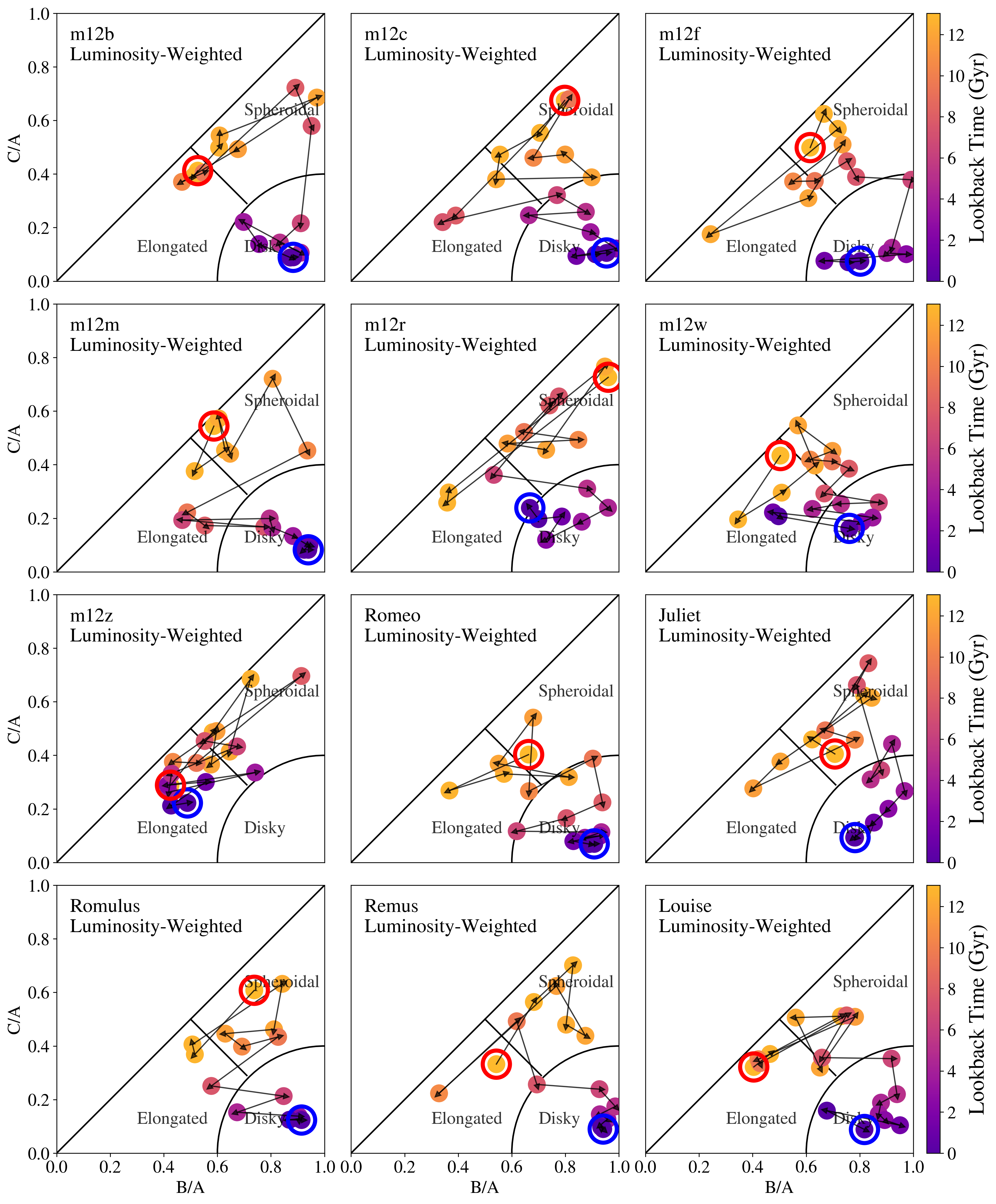}
    \caption{The luminosity-weighted shape evolution of simulated galaxies, corresponding to the bottom right panel of Figure \ref{fig:Thelmapoints}.}
    \label{allm12Lumweight}
\end{figure*}

\begin{figure*}
    \centering
    \includegraphics[width=\textwidth]{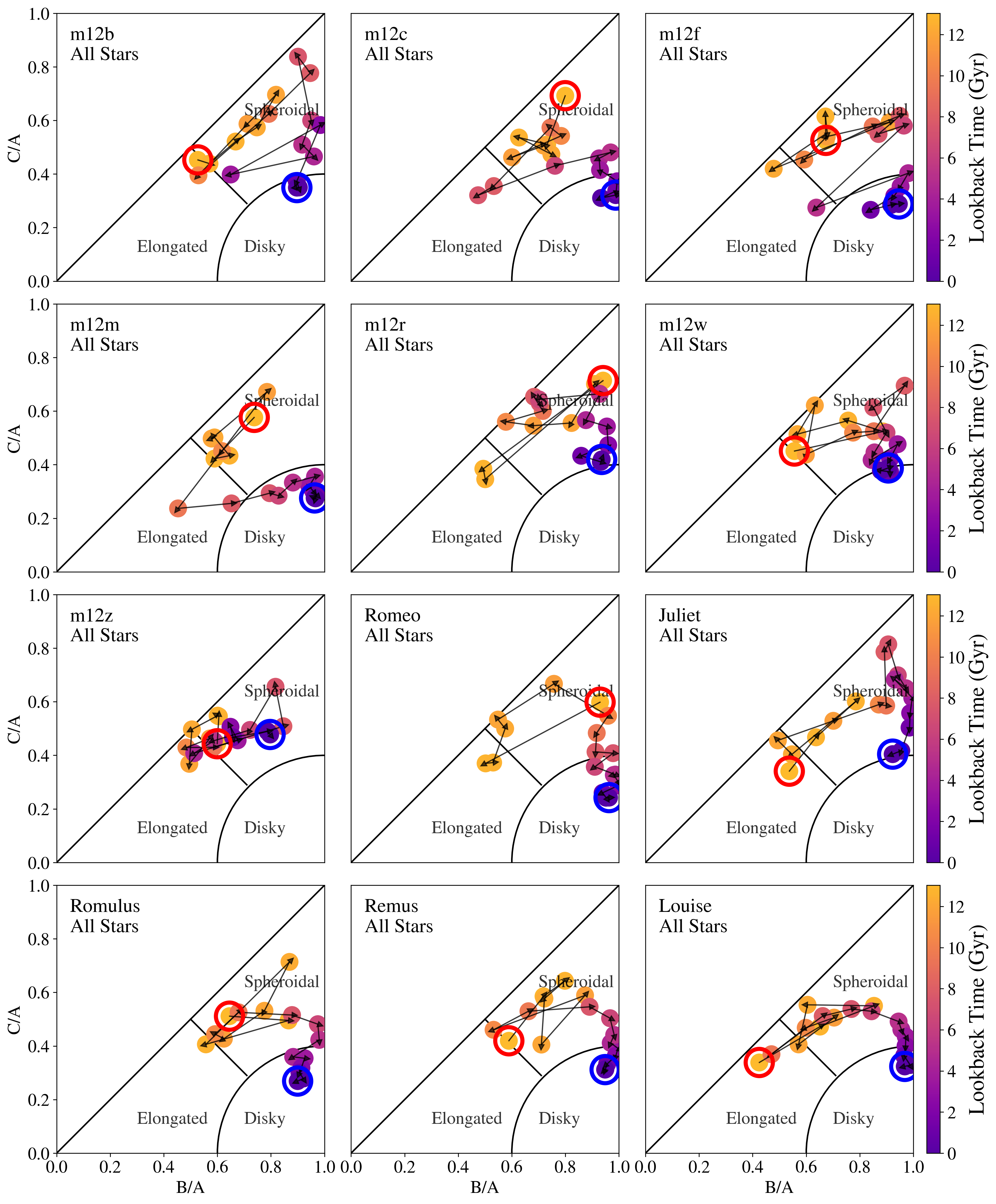}
    \caption{The all-stars shape evolution of the sample, corresponding to the bottom left panel of Figure \ref{fig:Thelmapoints}. Note that m12z has experienced a recent merger. It is the only galaxy that does not form a disk.}
    \label{allm12AllStars}
\end{figure*}

This appendix shows the shape evolution of the other 12 Milky Way--like galaxies besides Thelma, which was shown in Figures \ref{fig:Thelma_pastpresent} and \ref{fig:Thelmapoints}. Figure \ref{allgalaxiespastpresent} shows how the shapes of stellar populations change from birth to the present day. The archaeological decomposition for each galaxy is shown in Figure \ref{allm12Archaeology}. The remaining figures show the shape evolution using young stars (Figure \ref{allm12YoungStars}), luminosity-weighted all stars (Figure \ref{allm12Lumweight}), and all stars (Figure \ref{allm12AllStars}), respectively.

\bsp	
\label{lastpage}
\end{document}